   \documentclass{aa}
   \usepackage{graphicx}
   
\begin{document}
   \title{The 3--D ionization structure of 
NGC~6818: a Planetary Nebula threatened by recombination
   \thanks{Based on observations made with ESO Telescopes at the La Silla 
Observatories, under programme ID 65.I-0524, and on  
observations made with the NASA/ESA Hubble Space Telescope, 
obtained from the data archive at the Space Telescope Institute 
(observing programs GO 7501 and GO 8773; P.I. Arsen Hajian). 
STScI is operated by the association of Universities for Research in   
Astronomy, Inc. under the NASA contract  NAS 5-26555.  We have
applied the photo--ionization code CLOUDY, developed at the Institute of
Astronomy of the Cambridge University.}
   \author{S. Benetti \inst{1} \and E. Cappellaro \inst{2} \and
  R. Ragazzoni \inst{3} \and F. Sabbadin  \inst{1} \and M. Turatto
   \inst{1}}
   \offprints{S. Benetti, benetti@pd.astro.it}
   \institute{INAF - Osservatorio Astronomico di Padova, vicolo dell'Osservatorio 5, I-35122 Padova, Italy \and 
   INAF - Osservatorio Astronomico di Capodimonte, via Moiariello 11, I-80131 Napoli, Italy \and INAF - Osservatorio 
   Astrofisico di Arcetri, Largo E. Fermi 5, I-50125, Italy}}
   \date{Received September 24, 2002; accepted November 11, 2002}
   
   \abstract{Long-slit NTT+EMMI echellograms of NGC 6818 (the
   Little Gem) at nine equally spaced position angles, reduced
   according to the 3-D methodology introduced by Sabbadin et
   al. (2000a,b), allowed us to derive: the expansion law, the
   diagnostics and ionic radial profiles, the distance and the central
   star parameters, the nebular photo-ionization model, the 3-D
   reconstruction in He II, [O III] and [N II], the multicolor
   projection and a series of movies. The Little Gem results to be a
   young (3500 years), optically thin (quasi--thin in some directions)
   double shell (M$_{\rm ion}$$\simeq$0.13 M$_\odot$) at a distance of 1.7
   kpc, seen almost equatorial on: a tenuous and patchy spherical
   envelope (r$\simeq$0.090 pc) encircles a dense and inhomogeneous
   tri-axial ellipsoid (a/2$\simeq$0.077 pc, a/b$\simeq$1.25,
   b/c$\simeq$1.15) characterized by a hole along the major axis and a
   pair of equatorial, thick moustaches. NGC 6818 is at the start of
   the recombination phase following the luminosity decline of the
   0.625 M$_\odot$ central star, which has recently exhausted the
   hydrogen shell nuclear burning and is rapidly moving toward the
   white dwarf domain (log T$_*\simeq$5.22 K; log
   L$_*$/L$_\odot\simeq$3.1).  The nebula is destined to become thicker
   and thicker, with an increasing fraction of neutral, dusty gas in
   the outermost layers. Only over some hundreds of years the plasma
   rarefaction due to the expansion will prevail against the slower and
   slower stellar decline, leading to a gradual re-growing of the
   ionization front.  The exciting star of NGC 6818 (m$_{\rm V}\simeq$17.06)
   is a visual binary: a faint, red companion (m$_{\rm V}\simeq$17.73)
   appears at 0.09 arcsec in PA=190$\degr$, corresponding to a
   separation $\ge$150 AU and to an orbital period $\ge$1500 years.
   }
   
   \titlerunning{The Planetary Nebula NGC 6818}
   
   \maketitle
%
\section{Introduction} 

It is widely accepted that an ``aged'' star of low to medium mass (1.0
M$_\odot$$<$M$_{\rm MS}$$<$8.0 M$_\odot$) which is evolving toward the white
dwarf region of the H--R diagram, first expels and then ionizes
the surface layers, thus generating the Planetary Nebula (PN)
phenomenology (Paczynski 1970, Aller 1984, Pottasch 1984, Osterbrock
1989).
\begin{figure*} \centering
\includegraphics[width=7cm]{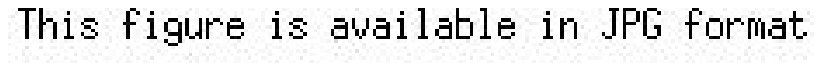} \caption{WFPC2 archive images 
of NGC~6818 in [O III] (left) and [N II] (right). The binary nature of the 
central star is shown in the enlargement at the bottom. North is up and East to the left.}  
\end{figure*}

Thanks to the advent of sophisticate photo-ionization codes
(Harrington 1989, Pequignot 1997, Ferland et al. 1998), we now know in
detail the physical effects produced by any UV flux on any gas
distribution and composition.  Conversely, the situation is quite
disappointing for a true nebula, due to projection limitations: on the
one hand the HST imagery has strongly enhanced the ground-based
evidences of the PNe complexity, on the other hand the observational
data are still interpreted in terms of approximate structures and
unrealistic assumptions for the physical conditions, like
$T{\rm e}$=constant and $N{\rm e}$=constant all over the object (more comments are
in Aller 1984, 1990, 1994).

In order to overcome the wide gap between theory and practice, the
apparent, bi-dimensional nebular image should be de-projected, and the
accurate spatial distribution of the gas recovered.

To this end we have developed an original procedure based on high
dispersion spectra: the PN being an extended and expanding plasma, the
position, thickness and density of each elementary volume can be, in
principle, obtained from the radial velocity, width and flux of the
corresponding emission.  We first apply a tomographic analysis, which
reconstructs the ionic distribution in the nebular slices covered by
the spectrograph slit, and then assemble all the tomographic maps by
means of a 3-D rendering procedure for studying the morphology,
physical conditions, ionization, spatial structure and evolutionary
status.

The rationale of the method and the earliest, rough results based on plate echellograms go back to 
Sabbadin et al. (1985, 1987). More recent, ``quantitative'' observations (i.e. using a linear detector) 
concern NGC 40 and NGC 1501, both objects covered at moderate spectral resolution, R$\simeq$ 20000--25000, 
with the Echelle + 1.82m telescope of Padua Observatory at Asiago, Cima Ekar (Sabbadin et al. 2000a, b, and 
Ragazzoni et al. 2001; hereafter Papers I to III, respectively).

At the same time we have carried out a survey of two dozen  
PNe and proto-PNe in both hemispheres with ESO NTT+EMMI (spectral range
$\lambda\lambda$3900-7900 $\AA\/$, R=60000, spatial resolution$\simeq$1.0 arcsec) and the Telescopio Nazionale 
Galileo (TNG)+SARG (spectral range  
$\lambda\lambda$4600-8000 $\AA\/$, R= 115000, spatial resolution$\simeq$0.7 arcsec). 
The observed sample covers a variety of morphologies, kinematics and evolutionary phases, including  
NGC 2392 (the Eskimo nebula), NGC 3132 (the Eight-Burst n.), NGC 3242 (the Ghost of Jupiter n.), NGC 6210 (the Turtle n.), 
NGC 6543 (the Cat's Eye n.), NGC 6751 (the Glowing Eye n.), 
NGC 6826 (the Blinking n.), NGC 7009 (the Saturn n.), 
NGC 7662 (the Blue Snowball n.), IC 418 (the Spirograph n.), He 2-47 (the Starfish n.), MyCn 18 (the Hourglass n.), 
MZ 3 (the Ant n.), and as many un-dubbed, but equally exciting targets.

The 3-D ionization structure of NGC 6565 has been presented by 
Turatto et al. (2002, Paper IV); here we discuss the case of NGC 6818 (PNG 025.8-17.9, Acker et al. 1992).

The paper is structured as follows: Sect. 2 introduces the nebula, Sect. 3 presents the 
observational 
material and the reduction procedure, Sect. 4 is dedicated to the 
gas kinematics, Sect. 5 concerns the radial profile of the physical conditions (electron 
temperature and electron density from forbidden line ratios), in Sect. 6  the ionization 
structure and the overall gas distribution are discussed, in Sect. 7 
we derive the nebular distance, mass and age, in Sect. 8 the central star parameters are given,
Sect. 9 contains the application of the photo-ionization model (CLOUDY),
Sect. 10 describes the 3-D structure of the nebula in different ions, and Sect. 11 presents a
short discussion and the conclusions.

\section{The nebula}
The HST/WFPC2 appearance of the high excitation PN NGC 6818 (sometimes called the Little Gem) is shown in Fig. 1: 
in [O III] ``a roughly 
spherical outer envelope as well as a brighter vase-shaped interior bubble. 
There is a prominent orifice to the North and a smaller one to the South, along 
the major axis, probably caused by a blow-out from a fast wind'' (Rubin et al. 
1998). A few dark filaments and knots are also present.

The [N II] emission mainly occurs in two irregular equatorial ``moustaches'' and in 
a multitude of condensations (in some cases presenting a sort of radially 
arranged cometary tail), which are brighter in the southern part of the nebula. 

The striking HST multicolor reproduction by Arsen Hajian \& Yervant
Terzian (http://ad.usno.navy.mil/pne/ gallery.html) highlights the
composite envelope structure and the large stratification of the
radiation within NGC 6818.

Rubin et at. (1998) reported the presence of two faint stars, roughly 2--4 arcsec N and NE of 
the central star, which could be physically associated to it. A careful analysis of the WFPC2 frames 
allowed us to unveil the true binary nature of the central star: the red, faint companion is at a distance 
of 0.093($\pm$0.004) arcsec in position angle (PA)= 190($\pm$2)$\degr$ (see the enlargement at the bottom of Fig. 1). 

Following Weedman (1968) our nebula is a prolate spheroid (an ellipse rotated about the major axis) 
of moderate ellipticity (axial ratio 3:2) observed perpendicular to the major axis, whereas 
Sabbadin (1984) suggests a somewhat optically thick prolate spheroid in a early evolutionary 
phase, seen almost pole-on. 
According to Hyung et al. (1999), a definitive statement on the geometry must wait for more adequate kinematical data, due 
to the complexity of NGC 6818. 

To this end we have secured a series of long-slit, radially arranged echellograms. They were analysed with the 
3-D methodology described in Papers I to IV.

\section{The observational material and the reduction technique}

\begin{figure*}
   \centering \includegraphics[width=7cm]{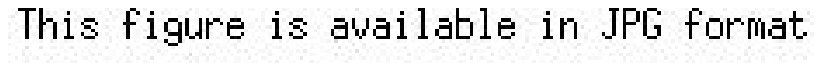}
   \caption{Detailed structure (in logarithmic scale) of some representative emissions
at the nine observed PA of NGC 6818. The blue-shifted gas is to the left. 
The slit orientation at each PA is indicated 
in the corresponding [O I] frame. The $\lambda$6300.304 $\AA\/$ night sky
line has been removed from the [O I] spectral image.}  
\end{figure*}

Nine echellograms of NGC~6818 (exposure time 600s; spectral
resolution 60000 with a slit 1.0 arcsec wide) have been obtained on
July 29, 2000 at the ESO NTT in photometric sky conditions and seeing between 0.7 and 1.0 arcsec. We have used a 
40 arcsec long slit centered on the exciting star for all the selected PA, 
ranging from 10$\degr$ to 170$\degr$ with a constant step of 20$\degr$. 
Since we do not insert an interference filter (as normally done by the other observers to isolate a single order),  
each spectrum covers 80 echelle orders ($\lambda$3967 $\AA\/$ of [Ne III] to $\lambda$7751 $\AA\/$ of [Ar III]), 
and provides the  
spatio--kinematical structure of a good two dozen nebular emissions, representing all the main ionic species.  

The reduction method follows conceptually the standard procedure,
including bias, flat field, distortion correction, wavelength and
flux calibration, and is carefully described in Paper IV.

Fig. 2 illustrates the detailed structure in [O I] ($\lambda$6300 $\AA\/$), [N II] ($\lambda$6584 $\AA\/$), H I 
($\lambda$6563 $\AA\/$), [O III] ($\lambda$5007 $\AA\/$) and He II ($\lambda$4686 $\AA\/$) at the 
observed PA. 

NGC 6818 exhibits a complex ionization structure: 
\begin{description} 
\item[-] [O I] is only seen 
in the outermost regions, in the form of distinct condensations, which in some cases are 
symmetrically arranged (the moustaches), otherwise single (like the cometary knot in PA=150$\degr$, 
southern edge); 
\item[-] the low ionization [N II] emission forms a double envelope of inhomogeneous 
structure from PA=50$\degr$ to PA=130$\degr$, and a single, elongated and distorted ring from PA=150$\degr$ to 
PA=30$\degr$. The presence of  FLIERS (fast, low ionization emitting regions), as introduced by 
Balick et al. (1993, see also Corradi et al. 1996), is not confirmed here: although some isolated spikes present a 
faint, high velocity 
tail, the overall [N II] emission of NGC 6818 can be understood in terms of ``normal'' 
nebular regions simply shadowed by some inner and dense layer causing the ionization drop in the 
outer plasma;  
\item[-] the H$\alpha$ line (central row in Fig. 2) is characterized by a very blurred appearance 
(due to a combination of thermal motions plus fine structure plus expansion velocity gradient), masking the 
detailed distribution of the ionized gas. The same effect is also present in He I and He II, but 
it is particularly damaging for H$\alpha$, the reference emission in both the radial electron density 
and ionization determinations (Paper IV);
\item[-] $\lambda$5007 $\AA\/$ of [O III] (a mean-high excitation ion), by far the strongest line in the 
optical region, highlights the double envelope structure of NGC 6818: the inner shell 
(better seen between PA=50$\degr$ and PA=130$\degr$) is an irregular ellipsoid broken along the 
major axis (N-S direction) whose equatorial, denser regions are identified by the ``moustaches''. 
The line--tilt between PA=90$\degr$ and  PA=110$\degr$ suggests that we are misaligned 
with both the intermediate and the minor axes of the inner ellipsoid, and that the line of the nodes is at 
PA$\simeq$60$\degr$. The outer shell is almost un-tilted, spherical, 
holed at North and South, and circumscribes the internal one; 
\item[-] the He II emission (bottom row in Fig. 2) marks the highest excitation nebular regions,  
mainly  constituted by the inner shell. Note the blurred appearance of $\lambda$4686 $\AA$, 
essentially due to the thirteen fine structure components.
\end{description}
The first, qualitative picture of the spatial structure coming from the echellograms confirms the 
indications already obtained from the imaging (Sect. 2).

\section{The gas kinematics}

According to Papers I to IV, 
the expansion velocity ($V{\rm exp}$) of the ionized gas can be
derived from the analysis of the ``central star pixel line'' (cspl) in the different ions. The cspl 
is parallel to the dispersion, selects the nebular material 
projected at the apparent position of the star (whose motion is purely radial) and is the same 
in all the frames, the slit being radially arranged. Thus, in order to improve the S/N of the faintest 
emissions, the nine echellograms have been combined.
\begin{table*}
\caption{Peak separation in the cspl of NGC~6818 }
\begin{tabular}{ccccccc}
\hline
\\
\\
Ion &IP (eV)&&&2$V{\rm exp}$ (km/s)&&\\
\cline {3-7}
&&Wilson 1950& Sabbadin 1984&Meatheringham et al. 1988&Hyung et at. 1999&this paper  \\
\\
$[$O I$]$   &  0.0  &     -    &      -    &     -    &    -  &72:  \\
$[$S II$]$  & 10.4  &     -    &      -    &     -    &   60.7&71  \\
$[$O II$]$  & 13.6  &    60.2  &      -    &    59.1  &   58.9&69:  \\
H I         & 13.6  &    55.5  &     52    &     -    &   42.6&55  \\
$[$N II$]$  & 14.5  &     -    &     64    &     -    &   61.8&70  \\
$[$S III$]$ & 23.4  &     -    &      -    &     -    &   50.8&65:  \\
He I        & 24.6  &     -    &      -    &     -    &   57.3&66  \\
$[$Ar III$]$& 27.6  &     -    &      -    &     -    &   48.9&64  \\
$[$O III$]$ & 35.1  &    56.2  &     54    &    55.1  &   51.3&62  \\
$[$Ar IV$]$ & 40.7  &     -    &      -    &     -    &   30.7&48  \\
$[$Ne III$]$& 41.0  &    58.0  &      -    &     -    &   52.2&63  \\
N III       & 47.4  &     -    &      -    &     -    &   41.5&-  \\
He II       & 54.4  &    42.4  &      -    &    41.9  &   31.5&45  \\
$[$Ar V$]$  & 59.8  &     -    &      -    &     -    &   20.2&36  \\
$[$Ne V$]$  & 97.1  &    32.6  &      -    &     -    &    -  &-  \\
\\
\hline
\end{tabular}
\end{table*}

The results are contained in the last column of Table 1, where the
ions are put in order of increasing ionization potential (IP). Typical
errors are 1.5 km s$^{-1}$ for the strongest forbidden emissions (like
$\lambda$6584 $\AA\/$ of [N II] and $\lambda$5007 $\AA\/$ of [O III])
to 3.0 km s$^{-1}$ for the faintest ones (in particular: $\lambda$6300
$\AA\/$ of [O I], because of the knotty structure, and
$\lambda$7319.87 $\AA\/$ of [O II], which is partially blended with
$\lambda$7318.79 $\AA\/$, also belonging to the O$^+$ red
quartet). The corresponding uncertainties for the recombination lines
are: 2.5 km s$^{-1}$ for $\lambda$6563 $\AA\/$ (H I), and 2.0 km
s$^{-1}$ for $\lambda$5876 $\AA\/$ (He I) and $\lambda$4686
$\AA\/$ (He II).

Columns 3 to 6 of Table 1 report the kinematical results from  
the literature. In detail: 
\begin{description}
\item[-] Wilson (1950) obtained a single Coud\`e spectrum of NGC 6818 (without de-rotator) 
at spectral resolution R$\simeq$30000;
\item[-] Sabbadin (1984) observed the nebula 
at four PA (long-slit echellograms at R$\simeq$15000);
\item[-] Meatheringham et al. (1988, 
long-slit echellograms at unspecified position, R$\simeq$26000) also measured 
the width at 10$\%$ maximum intensity of $\lambda$5007 $\AA\/$, obtaining 2$V{\rm exp}$[O III]= 87.6 km s$^{-1}$ 
(following Dopita et al. 1985, this corresponds to the largest expansion velocity of the gas); 
\item[-] Hyung et al. (1999, R$\simeq$33000) studied 
four bright regions located along the apparent minor and major axes (4--5 arcsec 
East and West, and 9--10 arcsec North and South of the central star); their
values represent lower limits to $V{\rm exp}$(various ions), due to projection effects. 
\end{description}

In Table 1 the kinematical variety reported by the different authors is symptomatic of the difficulties 
connected to the analysis of 
the high dispersion spectra, and, according to Hyung et al. (1999), stresses the fact that only the detailed coverage at 
adequate spatial and spectral resolutions can provide a reliable information on the structure of a chaotic 
object like NGC 6818. 

As cspl refers to the kinematical properties of the 
matter projected at the apparent position of the star, so the ``zero velocity pixel column'' (zvpc), 
corresponding to the recession 
velocity of the whole nebula, gives the radial distribution of the ionized gas which is 
expanding perpendicularly to the line of sight (Paper IV and references therein).

The intensity peak separations, 2r$_{\rm zvpc}$, 
in the different emissions at the nine PA of NGC 6818 are presented in Table 2. 
Both Fig. 2 and Table 2 indicate complex zvpc profiles, often multi-peaked (inner and outer shell), fast 
changing in direction, and variable (at a given PA) from ion to ion. The observed line structure  
depends on the radial matter distribution, the ionization and the intensity of the emission (the same occurs 
for the cspl assuming $V{\rm exp}$$\propto$r, as normally observed in PNe).

To be noticed: the ``relative'' spectral and spatial resolutions of our echellograms, as introduced  
in Paper III, differ by a factor of two. The former is   
given by RR=$V{\rm exp}$/$\Delta$V $\simeq$ 6, $\Delta$V being the spectral resolution, and the latter by 
SS=r/$\Delta$r $\simeq$ 12 (r=apparent radius, $\Delta$r=seeing). In practice this means that the 
spatial information of NGC 6818 is twice as detailed as the kinematical one.
\begin{centering}
\begin{table*}
\caption{Peak separation in the zvpc at the nine observed PA of NGC~6818 }
\begin{tabular}{cccccccccc}
\hline
\\
Ion&&&&&2r$_{\rm zvpc}$&(arcsec)&&&\\ 
\cline{2-10}
&PA=10$\degr$ &PA=30$\degr$ &PA=50$\degr$ &PA=70$\degr$ &PA=90$\degr$ &PA=110$\degr$ 
&PA=130$\degr$ &PA=150$\degr$&PA=170$\degr$  \\
\\
$[$O I$]$     & 23.5:  & 23.0: & 22.0: &  ---;  22.5: &---;  22.5: &---;  22.5:  &---;  22.0:  &---;  22.5:  &  22.3: \\
$[$S II$]$    & 23.5   & 22.4  & 22.0  & 15.6;  22.4  &---;  22.3: &15.8;  22.5 &14.0;  22.0 &---;  22.4   &  22.6  \\
$[$O II$]$    & 23.1:  & 22.8: & 21.7: &  ---;  22.2: &---; 21.7: &---;  22.5:  &---;  22.0:  &---;  22.0:  &  21.8: \\
HI           & 19.2:  & 16.7:  & 16.4: & 14.0:; 21.0: &12.8:; 19.5:&13.0:; 21.0: & 13.1:;  20.5: &14.7:;  21.0: &  20.2:  \\
$[$N II$]$    & 23.5   & 22.4  & 22.0  & 15.8;  22.4  &---;  22.2 &15.6;  22.4 & 14.5;  22.4 &16.0;  22.4 &  21.8  \\
$[$S III$]$   & 21.3   & 20.5  & 18.4: & 15.3;  ---    &15.5:; --- & 14.5:; ---& 13.4;  21.3:& 15.0;  20.5:& 19.9: \\
He I          & 22.2   & 20.5  & 18.1  & 15.7;  22.2  & 14.5;  20.6& 14.5:; 21.0 & 13.2:;  21.3&---;  19.2:   &  20.8  \\
$[$Ar III$]$  & 22.5   & 19.8  & 17.6  & 15.4;  22.2  & 14.6;  20.5& 14.7;  21.0 & 13.7;  21.4 & 15.4;  21.9 &  21.1  \\
$[$O III$]$   & 20.7   & 20.0  & 18.5  & 15.3;  21.2  & 15.0;  20.2& 15.0;  21.0 & 13.8;  20.0 & 15.9;  21.0 &  20.4  \\
$[$Ar IV$]$   & 17.9:  & 15.3  & 13.1: & 12.3;  ---    & 11.0;  ---  & 11.1;  ---   & 12.0;  --- &  12.7;  ---   &  16.4  \\
$[$Ne III$]$  & 21.7:  & 20.9: & 20.0: & 15.3:; ---   & 15.2:; 21.0:& 14.5;  21.0& 14.3;  20.4 & ---;  21.3:  &  21.6  \\
He II         & 19.0   & 15.2  & 12.3  & 11.8;  ---   & 11.0;  ---  &  11.0;  ---  & 12.1;  ---  &12.6;  --- &  16.6  \\ 
$[$Ar V$]$    & 17.7:  & 13.3: & 10.6: &  9.8;  ---   & 8.0:; ---   &  8.5;  ---   &  9.9;  --- & 10.9;  ---  &  13.7  \\

\\
\hline
\end{tabular}
\end{table*}
\end{centering}

In order to assemble the kinematical and the spatial results (corresponding to a single radial direction and to nine tangential 
directions, respectively) we must identify the PA at which r$_{\rm zvpc}$$\simeq$r$_{\rm cspl}$. 
The solution comes from the qualitative picture resulting from 
both the imaging (Sect. 2) and the spectra (Sects. 3 and 4). Let's consider for the main component of NGC 6818 the most 
general spatial structure, i. e. a tri-axial ellipsoid. The major axis, projected in PA$\simeq$10$\degr$, is almost 
perpendicular to the line of sight, since the emissions, although inhomogeneous and distorted by irregular motions, 
appear un-tilted. The intermediate and the minor axes lie in PA$\simeq$100$\degr$, and the line-tilt observed at PA=90$\degr$ 
and 110$\degr$ indicates that we are misaligned with both these axes. Thus, r$_{\rm zvpc}$$\simeq$r$_{\rm cspl}$ 
along the apparent minor axis of 
the nebula, that is between PA=90$\degr$ and 110$\degr$.
 
The cspl--zvpc connection for the inner and the outer shells is illustrated in Fig. 3, showing both
r$_{\rm zvpc}$ (at PA=90$\degr$ and 110$\degr$) and  $V{\rm exp}$ vs IP. For reasons of 
homogeneity each zvpc peak separation is normalized to the corresponding [O III] value. 

\begin{figure*} \centering
\includegraphics[angle=-90,width=14cm]{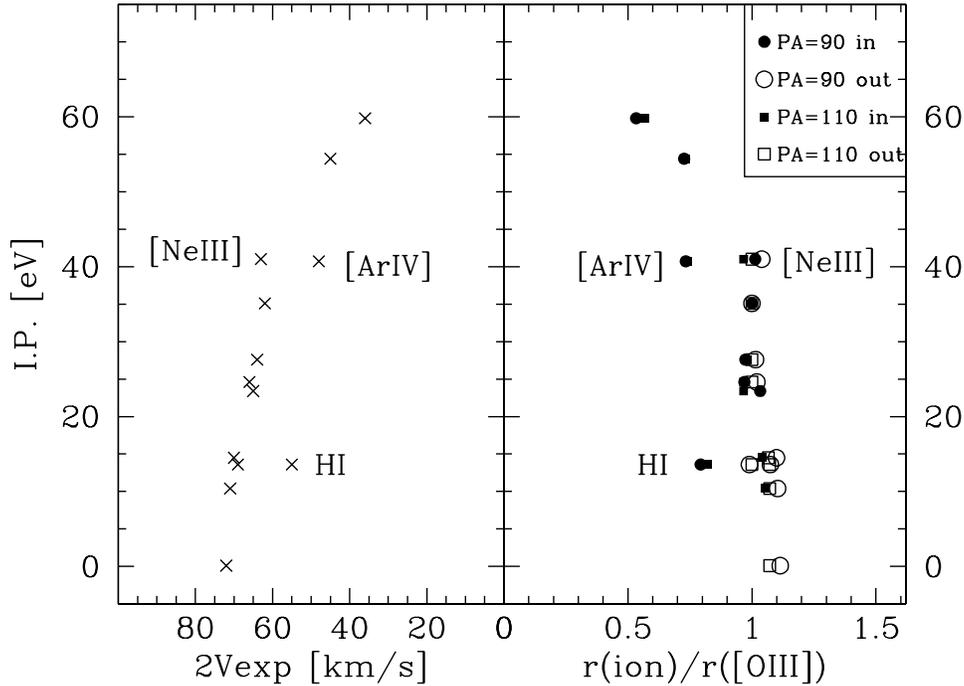} \caption{Expansion velocity in the cspl (left) and peak separation 
in the zvpc for the inner and the outer shells of NGC 6818 at PA=90$\degr$ and 110$\degr$ (right) vs IP. Each zvpc peak 
separation is normalized to the corresponding [O III] value. The ions in peculiar positions are marked.}  
\end{figure*}

NGC 6818 is characterized by a large stratification of the radiation and of the kinematics, in agreement with 
the Wilson's (1950) law. However, $V{\rm exp}$ and r$_{\rm zvpc}$ of [Ar IV] and [Ne III] 
considerably deviate from the sequence defined by the other ionic species. 
This behavior, also noticed 
in NGC 6565 (Paper IV), is typical of a PN powered by a high temperature central star (for 
details, see the above-mentioned reference and Sect. 9). 
Instead, the anomalous position of H I in Fig. 3 is the obvious consequence of the mono-electron atomic structure 
of hydrogen. 

The r$_{\rm zvpc}$ vs IP relation being the same at all the observed PA
(from Table 2), we can assess that in NGC 6818 the expansion
velocity is proportional to the distance from the central star through
the relation:

\begin{equation}
  V{\rm exp}(km s^{-1}) = 3.5 (\pm0.3)\times r'' 
\end{equation}

Weedman (1968) has derived $V{\rm exp}(km s^{-1})$ =$5.9\times(r''-0.8)$,
which is almost twice as steep as Eq. (1). The Weedman's expansion law
appears questionable, since it is based on the [O III] profile in a
single, quite under-exposed Coud\`e spectrum (R$\simeq$60000) taken
along the apparent major axis (PA$\simeq$10$\degr$). Weedman noticed
that $\lambda$5007 $\AA$, although distorted by irregular motions, is
un-tilted, and assumed ``a priori'' the following model: a prolate
spheroid with a=17.3 arcsec and a/b=1.5, seen perpendicular to the
major axis.
 
\begin{figure*} \centering
\includegraphics[width=7cm]{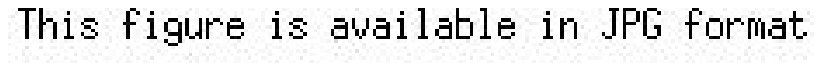} 
\caption{Position--velocity maps at the nine observed PA for the high (He II, blue), medium ([O III], 
green) and low ([N II], red) ionization regions of NGC 6818, scaled according to the relation 
$V{\rm exp}$(km s$^{-1}$)=
3.5$\times$r''. The orientation of these tomographic maps is the same of Fig. 2.}  
\end{figure*}

Therefore in the following we will consider Eq. (1) as representative
of the whole nebular kinematics.

All this is synthesized in Fig. 4, showing the position--velocity
(P--V) maps, i.e. the complete radial velocity field at the nine
observed PA.  They are relative to the systemic heliocentric velocity
of the nebula, Vr$_{\odot}$=-14.5($\pm1.0$) km s$^{-1}$, corresponding
to V$_{\rm LSR}$=-1.7($\pm$1.0) km s$^{-1}$, and are scaled according to
Eq. (1). In other words, they reproduce the tomographic maps in the
nebular slices covered by the slit.  We have selected He II, [O III]
and [N II] as representative of the high, medium and low ionization
regions, respectively.

These multicolor P--V maps highlight the large stratification of the radiation and the chaotic nebular structure. 
NGC 6818 is optically thin in most directions. The low excitation [N II] emission is associated to the 
presence of some inner and dense layer (see the moustaches, for example). The outermost, faint [N II] spikes detected  
in PA=10$\degr$, 90$\degr$, 150$\degr$ and 170$\degr$ belong to the external shell.

The existence of a prominent hole to the North and of a smaller one to the South, as suggested by Rubin et al. 
(1998), is confirmed. The southern cavity is less evident in Fig. 4, being only grazed by our spectra at 
PA=170$\degr$ and 10$\degr$. Further hollows are present (like at PA=70$\degr$ and 90$\degr$, in the approaching gas at West 
of the central star position). 

The radial ionization structure of the S-SE edge at PA=150$\degr$,
corresponding to the region of the ``cometary knot'' visible in
Fig. 1, is puzzling: [O III] is lacking, whereas He II and [N II] are
intense in the inner and the outer layers, respectively.

In summary: although some spectral features of the northern and southern holes can be tentatively interpreted in terms 
of ionized gas moderately accelerated by some blowing-up agent (as suggested by Rubin et al. 1998), the linear 
expansion law here adopted represents a valid approximation of the overall nebular kinematics.

\section{The physical conditions}

\subsection{General considerations}

We can obtain the $T{\rm e}$ radial profile from the diagnostic line ratios 
of ions in p$^2$ or p$^4$ configurations (like [O III] and [N II]), and the $N{\rm e}$ radial distribution from both
the diagnostics of ions in p$^3$ configuration (like [S II]) and the absolute H$\alpha$ flux. According to 
Paper IV, the zvpc, which is independent on the expansion velocity field, must be used. 

In the specific case of NGC 6818 the large H$\alpha$ broadening 
(due to thermal motions, fine structure and expansion velocity gradient) prevents the accurate determination of 
F(H$\alpha$)$_{\rm zvpc}$ (and then of $N{\rm e}$(H$\alpha$)). 
Thus, in this Section $T{\rm e}$[O III], $T{\rm e}$[N II] and  $N{\rm e}$[S II] are derived from the corresponding line intensity 
ratios. Later on (Sect. 6.3) we will illustrate the adopted escamotage providing  
F(H$\alpha$)$_{\rm zvpc}$ and $N{\rm e}$(H$\alpha$) from the observed radial ionization structure and the assumption O/H=
constant across the nebula.

\subsection{Interstellar absorption}

First of all the observed line intensities must be corrected for interstellar absorption according 
to:

\begin{equation}
\log \frac{{\rm I}(\lambda)_{\rm corr}}{{\rm I}(\lambda)_{\rm obs}}=f_{\lambda}\, c({\rm H}\beta)
\end{equation}
where f$_{\lambda}$ is the interstellar extinction coefficient given by Seaton (1979). 
The logarithmic extinction at H$\beta$, c(H$\beta$), is normally obtained by comparing the
observed Balmer decrement (in particular H$\alpha$/H$\beta$) to the
intrinsic value given by Brocklehurst (1971) and Hummer \& Storey
(1987). The estimates of c(H$\beta$) (from the Balmer ratio) reported in the literature for NGC 6818 span the range 
0.25 (Aller \& Czyzak 1983, 
Liu \& Danziger 1993) to 0.41 (Collins et al. 1961). Moreover, Tylenda et al. (1992) and Condon et al. (1999) 
obtained c(H$\beta$)=0.33 and 0.40, respectively, from the radio to H$\beta$ fluxes. 

Thanks to the excellent spatial and spectral accuracies achieved by the
superposition technique used ($\pm0.15$ arcsec and $\pm1.0$ km s$^{-1}$,
respectively) we can extend the H$\alpha$/H$\beta$ analysis to
the whole spectral image, as recently introduced (Paper IV) in the study of  
NGC 6565, a compact, dust embedded  PN exhibiting a complex c(H$\beta$) profile. 
The results are less dramatic for NGC 6818, since the blurred appearance of both  H$\alpha$ and  
H$\beta$ limits the resolution: besides some indications of a soft decline in the innermost regions (likely caused 
by  a local increase of $T{\rm e}$), the spectral maps appear quite homogeneous at  H$\alpha$/H$\beta$=3.73 ($\pm$0.06), 
corresponding to  c(H$\beta$)=0.37 ($\pm$0.03) (for the case B of  Baker \& Menzel 1938, $T{\rm e}$=12000 K and 
log~$N{\rm e}$= 3.00; Brocklehurst 1971, Aller 1984, Hummer \& Storey 1987).

\subsection{$T{\rm e}$[O III], $T{\rm e}$[N II] and $N{\rm e}$[S II]}

First we obtain $T{\rm e}$[O III] from $\lambda$5007 $\AA$/$\lambda$4363 $\AA$, the line ratio being almost 
independent on $N{\rm e}$ for $N{\rm e}$$<$10$^4$ cm$^{-3}$.
$N{\rm e}$[S II] is then derived from  $\lambda$6717 $\AA$/$\lambda$6731 $\AA$ (using $T{\rm e}$[O III] to take into 
account the weak dependence of the ratio on $T{\rm e}$). 
Last, $T{\rm e}$[N II] comes from $\lambda$6584 $\AA$/$\lambda$5755 $\AA$ (adopting $N{\rm e}$[S II] for its weak dependence 
on the electron density).

Although the resulting $T{\rm e}$[O III], $T{\rm e}$[N II] and $N{\rm e}$[S II] profiles rapidly change with PA,  
as expected of the chaotic structure of NGC 6818, nevertheless there are some common features. 
In order to highlight both the differences and the analogies we have selected two PA close to the apparent 
major axis (PA=10$\degr$ and PA=30$\degr$) and two PA close to the apparent minor axis 
(PA=90$\degr$ and PA=110$\degr$) as representative of the whole nebular phenomenology.

\begin{figure} 
\centering
\includegraphics[width=9cm]{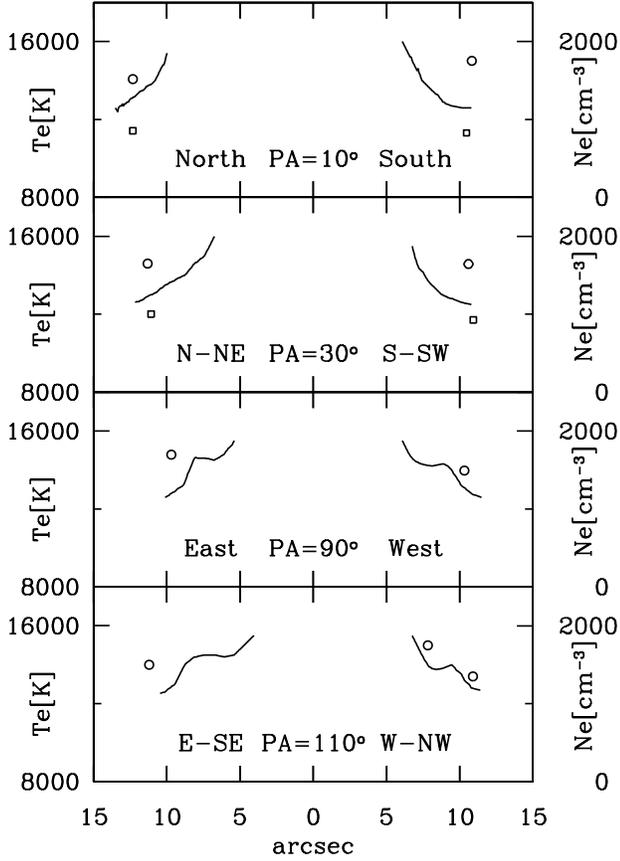} 
\caption{The diagnostics radial profile at PA=10$\degr$ and 30$\degr$ (close to the major axis of NGC 6818), and at 
PA=90$\degr$ and 110$\degr$ (close to the minor axis). Left ordinate scale: $T{\rm e}$[O III] (continuous line) and 
$T{\rm e}$[N II] (squares). Right ordinate scale: $N{\rm e}$[S II] (circles). $T{\rm e}$[N II] is lacking at PA=90$\degr$ and 110$\degr$, 
because of the auroral line weakness.}
  \end{figure}

The results are shown in Fig. 5; their main limitation is evident: due to the weakness of the [N II] auroral 
line and of the [S II] doublet, $T{\rm e}$[N II] e $N{\rm e}$[S II] can be obtained only at the 
corresponding intensity peaks (in the best case).

$T{\rm e}$[O III] presents a well-defined radial profile common to all the PA: it is $\ge$15000 K in the 
tenuous, innermost regions, it gradually decreases outward down to 12000--12500 K in the densest layers, and 
later it remains more or less constant (the last statement is weakened by the auroral line faintness). Such a radial 
trend is in quantitative agreement with the results by Rubin et al. (1998), based on $\lambda$4363 $\AA\/$ 
and $\lambda$5007 $\AA\/$ HST/WFPC2 imagery. The presence of a $T{\rm e}$ gradient across 
the nebula is also suggested by Hyung et al. (1999). Previous ground--based $T{\rm e}$[O III] determinations are mean values 
for the brightest (i.e. densest) regions and span the range 11400 K (de Freitas Pacheco et al. 1991) to 
12770 K (Mathis et al. 1998). 

In Fig. 5 $T{\rm e}$[N II] refers to the intensity peaks of the low ionization regions and is systematically below $T{\rm e}$[O III], 
in agreement with both the previous reports for NGC 6818 (9500 K, Hyung et al. 1999, to 11280 K, McKenna et al. 1996)  
and the general results for PNe (see Aller 1990, Gruenwald \& Viegas 1995, Mathis et al. 1998).  
More $T{\rm e}$ values for NGC 6818 are: 11500 K (C III], Kaler 1986), 12700 K and 13800 K ([Ne V] and [O IV], 
respectively, Rowlands et al. 1989), 14700 K (Balmer discontinuity, Liu \& Danziger 1993), 12500 K (C III], Mathis et 
al. 1998), 13000 K ([Cl IV], Hyung et al. 1999)  and 19000 K ([O II], Keenan et al. 1999).

In summary, the mean kinetic energy of the free electrons in NGC 6818 is quite large. This on the one hand 
explains both the H$\alpha$ broadening and the strength of the forbidden lines (in particular, $\lambda$5007 $\AA$ of 
[O III]), on the other hand is indicative of a very hot central star 
(T$_*$$\ge$ 150000 K, also supported by the presence of high excitation emissions, up to [Ne V], 
IP=97.1 eV; see Table 1).

Concerning the [S II] electron densities in the zvpc (Fig. 5), they are  limited to the brightest parts of the external, 
low ionization layers and show peaks up to 2000($\pm$200) cm$^{-3}$. 
Previously Hyung et al. (1999) obtained $N{\rm e}$[S II]$\simeq$ 2000 cm$^{-3}$ (they also report $N{\rm e}$[S II]$\simeq$3000 
cm$^{-3}$ from the improved calculations for 3 equivalent p--electrons by Keenan et al. 1996).

Besides the zvpc, we have extended the $\lambda$6717 $\AA$/$\lambda$6731 $\AA$ analysis to the prominent 
knots of the entire [S II] spectral images. The resulting $N{\rm e}$[S II] values span the range 1500($\pm$200) to 2800($\pm$200) 
cm$^{-3}$, the equatorial moustaches being the densest regions of NGC 6818.

In the next Section we will derive the local filling factor,
$\epsilon_{\rm l}$, by combining $N{\rm e}$[S II] and $N{\rm e}$(H$\alpha)_{\rm zvpc}$,
since $N{\rm e}$[S II]$\times$
$\epsilon_{\rm l}^{0.5}$$\simeq$N{\rm e}(H$\alpha)_{\rm zvpc}$ (Aller 1984,
Osterbrock 1989).


\section{The radial ionization structure}
  
\subsection{General considerations}

All the main ionic species are present in the echellograms, with the usual, dreadful handicap for hydrogen: 
due to the large H$\alpha$ broadening, the deconvolution for 
instrumental resolution plus thermal motions plus fine structure appears inadequate, and 
the detailed F(H$\alpha)_{\rm zvpc}$ profile cannot be determined. Moreover:

- the principal O$^+$ emissions 
($\lambda$3726 $\AA\/$ and $\lambda$3729 $\AA$) fall outside our
spectral range. We have considered the much weaker $\lambda$7319.87 $\AA\/$, 
the strongest line of the red O$^+$ quartet; 

-  $\lambda$3967 $\AA\/$ of [Ne III] being  at the extreme blue edge
of the frame, the quantitative Ne$^{++}$ analysis appears quite uncertain.

In absence of H$\alpha$ we are forced to adopt $\lambda$5007 $\AA\/$ as reference line, thus 
obtaining the radial ionization structure relative to O$^{++}$ according to: 
\begin{equation}
\frac{X^{+a}}{O^{++}} = \frac{F(\lambda(X^{+b}))_{\rm zvpc}}{F(\lambda 5007 \AA)_{\rm zvpc}} f(T{\rm e},N{\rm e})
\end{equation}
where a=b for the forbidden lines, and a=b+1 for the recombination ones.
Note that the solution of the equations of statistical equilibrium requires the detailed knowledge of the plasma 
diagnostics, whereas $N{\rm e}$[S II] has been derived only at the intensity peaks of the outermost, low ionization layers. 
Luckily, in the $N{\rm e}$ range here considered $f(T{\rm e},N{\rm e})$ reduces to $f(T{\rm e})$ for most ionic species. Moreover all 
the emissions, with the exception of $\lambda$5876 $\AA$ of He I and $\lambda$4686 $\AA$ of He II, are forbidden lines, 
whose emissivity is a direct function of $T{\rm e}$. Thus, in first approximation we can put $N{\rm e}$=1.5$\times$10$^3$ cm$^{-3}$, 
constant across the nebula; this introduces uncertainties up to $\pm$10$\%$ (for O$^+$/O$^{++}$ and Ar$^{+4}$/O$^{++}$),   
which do not modify the results here obtained.

\begin{figure*} \centering
\includegraphics[width=16cm]{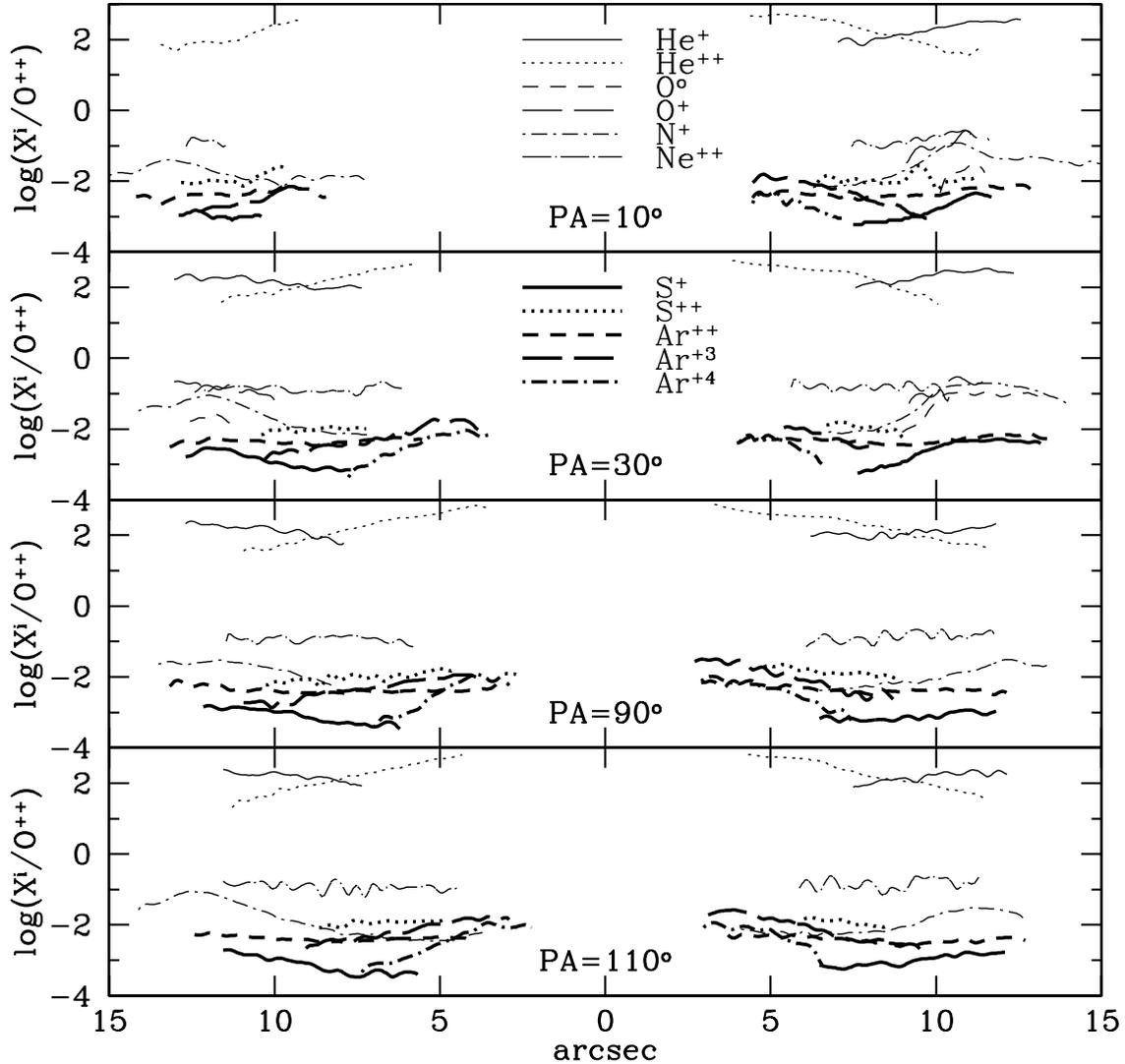} 
\caption{The radial ionization structure (relative to O$^{++}$) at PA=10$\degr$ 
and 30$\degr$ (close to the major axis of NGC 6818), and at PA=90$\degr$ and 110$\degr$ (close to the minor axis). The 
orientation is as in Fig. 5.}  
\end{figure*}

The radial profiles of $\frac{X^{+i}}{O^{++}}$ at the four selected PA of NGC 6818, shown in Fig. 6, contain 
a number of interesting features:
\begin{description}
\item[-] the weakness of the low excitation emissions in PA=10$\degr$ northern sector, PA=90$\degr$ both sectors and 
PA=110$\degr$ both sectors indicates that the nebula is optically thin in these directions. It is almost thick in PA=10$\degr$ 
southern sector and PA=30$\degr$ both sectors;
\item[-] as expected, He$^{++}$/O$^{++}$ decreases outward and He$^+$/ O$^{++}$ increases. They cross for
He$^{++}$/O$^{++}$=He$^+$/O$^{++}$ $\simeq$120, implying that He$_{\rm tot}$/O$^{++}\simeq$240. Since in these internal regions 
O$_{\rm tot}\simeq$1.30$\times$O$^{++}$ (see Eqs. (10) and (11) in Sect. 6.3) we derive He/O$\simeq$185;
\item[-] from similar considerations we infer N/O$\simeq$ 0.23, $N{\rm e}$/O$\simeq$ 0.15, 
S/O$\simeq$ 0.012  and Ar/O$\simeq$ 0.008.
\end{description}

These chemical abundances (relative to oxygen) must be compared with the corresponding values (relative to hydrogen) obtained 
from the conventional method, as illustrated in the next Section.

\subsection{Total chemical abundances}

According to the critical analysis by Alexander \& Balick (1997) we consider the total line fluxes (i.e. integrated over 
the whole spatial profile and the expansion velocity field). The resulting 
ionic abundances must be multiplied for the corresponding ICFs, the correcting
factors for the unobserved ionic stages. These were obtained both empirically (Barker
1983, 1986) and from interpolation of theoretical nebular models (Shields
et al. 1981, Aller \& Czyzak 1983, Aller 1984, Osterbrock 1989).

The final mean chemical abundances of NGC 6818, presented in Table 3 (last column), are in reasonable 
agreement with the previous 
estimates reported in the literature (also listed in the Table), and in excellent agreement with the 
indications of Sect. 6.1.

\begin{centering}
\begin{table*}
\caption{Total chemical abundances (relative to hydrogen)}
\begin{tabular}{cccccc}
\hline
\\
Element&Aller \& Czyzak & de Freitas Pacheco  et al. &Liu \& Danziger & Hyung et al. &This paper\\
 & (1983)&(1991)&(1993)&(1999)&\\
\\
He &   0.107                & 0.126               &0.114&  0.105&0.106($\pm$0.003) \\
C  &    4.47$\times$10$^{-4}$&   -                 &-&  8.0$\times$10$^{-4}$ &  - \\
N  &    1.41$\times$10$^{-4}$& 1.1$\times$10$^{-4}$&-&  4.0$\times$10$^{-4}$&1.4($\pm$0.2)$\times$10$^{-4}$\\
O  &    5.50$\times$10$^{-4}$& 5.25$\times$10$^{-4}$&6.37$\times$10$^{-4}$& 7.0$\times$10$^{-4}$& 5.5($\pm$0.5)$\times$10$^{-4}$\\
Ne &    1.23$\times$10$^{-4}$&   -                 &-&  1.0$\times$10$^{-4}$&9.0($\pm$2.0)$\times$10$^{-5}$\\
Na &    3.09$\times$10$^{-6}$&   -                 &-&  3.0$\times$10$^{-6}$& -\\
Mg &       -                   &   -                 &-&  3.0$\times$10$^{-5}$& -\\
Si &      -                   &   -                 &-&  9.0$\times$10$^{-6}$& -\\
S  &   8.9$\times$10$^{-6}$ & 1.2$\times$10$^{-5}$&-&  7.0$\times$10$^{-6}$&6.2($\pm$1.4)$\times$10$^{-6}$\\
Cl &    1.9$\times$10$^{-7}$ &   -                 &-&  3.0$\times$10$^{-7}$& -\\
Ar &   3.8$\times$10$^{-6}$ &   -                 &-&  4.0$\times$10$^{-6}$& 4.0($\pm$0.8)$\times$10$^{-6}$\\
K  &     1.02$\times$10$^{-7}$&   -                 &-&  2.0$\times$10$^{-7}$&-\\
Ca &     1.15$\times$10$^{-7}$&   -                 &-&  1.5$\times$10$^{-7}$&-\\
\\
\hline
\end{tabular}
\end{table*}
\end{centering}
\subsection{F(H$\alpha$)$_{\rm zvpc}$ and $N{\rm e}$(H$\alpha$)}

 As emphasized in Paper IV, the H$\alpha$ flux distribution in the zvpc, 
F(H$\alpha$)$_{\rm zvpc}$, is the fundamental   
parameter linking $N{\rm e}$ with both the spatial and the 
kinematical properties of the expanding plasma through the relation:

\begin{equation}
N{\rm e} = \frac{1.19\times 10^9}{T{\rm e}^{-0.47}} \times 
(\frac{F(H\alpha)_{\rm zvpc}}{\epsilon_{\rm l} \times r_{\rm cspl} \times D})^{1/2}
\end{equation}
where:
\begin{description}
\item[-] D is the nebular distance;
\item[-] r$_{\rm cspl}$ is the angular radius of the cspl (i.e. the nebular size in the 
radial direction);
\item[-] $\epsilon_{\rm l}$ is the ``local filling factor'', representing the fraction of the local 
volume actually filled by matter with density $N{\rm e}$. 
\end{description}

In order to recover F(H$\alpha$)$_{\rm zvpc}$ we start assuming O/H= 5.5$\times$10$^{-4}$, constant across the nebula.
At each radial position:

\begin{equation}
\frac{O}{H} =\frac{\sum_{i=0}^8 O^{0+i}}{H^0 + H^+}
\end{equation}

It can be written in the form: 
\begin{equation}
\frac{O}{H} =\frac{O^{++}}{H^+}\times icf(O^{++}) 
\end{equation}

That is:
\begin{equation}
\frac{O}{H} =\frac{F(\lambda 5007 \AA)_{\rm zvpc}}{F(H\alpha)_{\rm zvpc}}\times f(T{\rm e},N{\rm e})\times icf(O^{++}) 
\end{equation}

Thus obtaining:
\\
$F(H\alpha)_{\rm zvpc}$ = 
\begin{equation}
=1.9\times10^3 F(\lambda 5007 \AA)_{\rm zvpc}\times f(T{\rm e},N{\rm e})\times icf(O^{++})
\end{equation}

Also in this case $f(T{\rm e},N{\rm e})$ essentially reduces to $f(T{\rm e})$, with opposite trends for the emissivity of the 
forbidden and the recombination line.

\begin{figure*} \centering
\includegraphics[width=15cm]{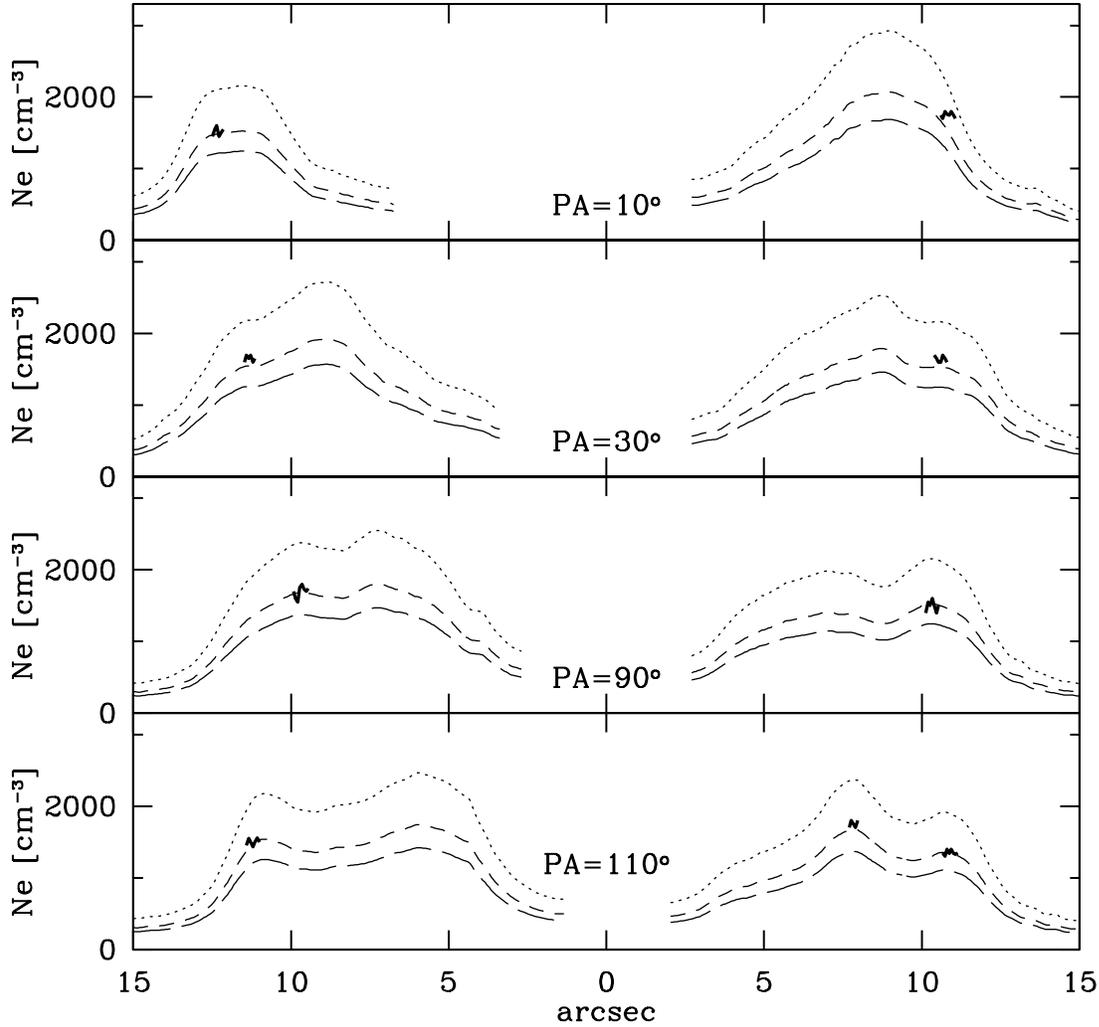} 
\caption{The $N{\rm e}$(H$\alpha)_{\rm zvpc}$ radial profile at the four selected PA of 
NGC 6818 for some representative values of $\epsilon_{\rm l}\times r_{\rm cspl}\times D$ (dotted line=5 arcsec kpc, 
short-dashed line=10 arcsec kpc, long--dashed line=15 arcsec kpc), superimposed to $N{\rm e}$[S II] 
(thick line; same as Fig. 5). The orientation is as in Figs. 5 and 6.}  
\end{figure*}

Eq. (8) provides the H$\alpha$ flux distribution in the zvpc (and
$N{\rm e}$(H$\alpha$) through Eq. (4)) once the ionization correcting factor
$icf(O^{++})$ is known. The complex structure of NGC 6818 implies that
$icf(O^{++})$ strongly changes across the nebula.  In the innermost
regions we have: O$^0$/O$^{++}$$<<$1, O$^+$/O$^{++}$$<<$1 and
H$^0$/H$^+$$<<$1.  Thus $icf(O^{++})$ becomes:

\begin{equation}
icf(O^{++})_{\rm inner} = 1 + \frac{\sum_{i=3}^8 O^{+i}}{O^{++}}
\end{equation} 

According to Seaton (1968), $\frac{\sum_{i=3}^8 O^{+i}}{O^{++}}$  can be derived from the ionization structure of 
helium thanks to the closeness of the O$^{++}$ and He$^+$ ionization potentials (54.9 and 54.4 eV, 
respectively). 
To this end we have performed a number of photo-ionization simulations 
(see also Alexander and Balick 1997) obtaining the fairly good relation:
\begin{equation}
\frac{\sum_{i=3}^8 O^{+i}}{O^{++}}\simeq 0.30\times\frac{He^{++}}{He^+} 
\end{equation}
 
Therefore we will adopt:
\begin{equation}
icf(O^{++})_{\rm inner} =  1+ 0.30\times\frac{He^{++}}{He^+} 
\end{equation}
 
Concerning the outermost nebula, the contribution of $\sum_{i=3}^8 O^{+i}$ can be neglected. 
Moreover in these regions we have O$^0$/O$^{+}$$<$1 (see Fig. 6), implying 
a low efficiency of the charge-exchange reaction 
O$^+$ + H$^0$$\getsto$O$^0$ + H$^+$ (Williams 1973; Aller 1984; Osterbrock 1989), i.e. H$^0$$<<$H$^+$. 
Thus:
\begin{equation}
icf(O^{++})_{\rm outer} =  1 + \frac{O^0}{O^{++}} + \frac{O^+}{O^{++}} 
\end{equation}

All this is summarized in Fig. 7, showing the $N{\rm e}$ radial profile (at the four selected PA of NGC 6818) obtained from 
F(H$\alpha$)$_{\rm zvpc}$ for some representative values of $\epsilon_{\rm l}\times$r$_{\rm cspl}$$\times$D (in arcsec
 kpc), superimposed to $N{\rm e}$[S II] (taken from Fig. 5). 

Fig. 7 evidences the basic advantage of using F(H$\alpha$)$_{\rm zvpc}$ in the determination of the electron density 
radial distribution: $N{\rm e}$(H$\alpha$) extends 
all over the nebular image, whereas $N{\rm e}$[S II] is limited to the peaks of the low excitation regions. 

In detail:
\begin{description}
\item[-] at PA=10$\degr$ (along the apparent major axis), $N{\rm e}$(H$\alpha$) presents a single, broad and asymmetric (i.e. steeper 
outwards) bell-shaped profile; 
\item[-] the double--peak structure is very subtle at PA=30$\degr$ (close to the apparent major axis), the 
inner peaks being predominant;
\item[-] at both PA=90$\degr$ and 110$\degr$ (close to the apparent minor axis) the two-shell 
distribution clearly appears; the $N{\rm e}$(H$\alpha$) top corresponds to the inner peaks at PA=110$\degr$, whereas at 
PA=90$\degr$ the peaks are almost equivalent;
\item[-] an outward, low density tail is present at all directions, extending up to about 15 arcsec from the star.
\end{description}

Note in Fig. 7 the match between $N{\rm e}$(H$\alpha$) and $N{\rm e}$[S II] for $\epsilon_{\rm l}\times$r$_{\rm cspl}$$\times$D$\simeq$ 
9.5($\pm$1) arcsec kpc, implying that $\epsilon_{\rm l}\times$D$\simeq$ 0.88 ($\pm$0.10) kpc  (see Table 2 and Sect. 4). 
Although the assumption $\epsilon_{\rm l}$=1 provides a lower limit of 0.9 kpc 
for the nebular distance, we can no longer delay a better quantification of this fundamental parameter.

\section{The nebular distance, mass and age}

The only individual distance reported in the literature (based on rough parameters for the star and the nebula) 
goes back to Gurzadyan (1970), who derived an indicative value of 2.2 kpc from the observed to expected size of 
the He$^{++}$ zone.

\begin{table*}
\caption{Statistical distances of NGC 6818}
\begin{tabular}{lll}
\hline
\\
Author & Distance (kpc)&Method      \\
\\  
O'Dell (1962) & 1.7& Shklovsky (1956), i.e. ionized mass=constant \\
Cahn \& Kaler (1971) &  1.82--2.38&Shklovsky \\
Cudworth (1974)& 1.95& proper motions\\
Milne \& Aller (1975)& 2.07& Shklovsky (radio)\\
Cahn (1976) &  2.2&Shklovsky\\
Acker (1978) & 1.6&published statistical distance scales re-calibrated with individual distances\\
Maciel \& Pottasch (1980)& 1.46 &ionized mass--radius relation\\
Daub (1982) & 1.29 & ionized mass--radio surface brightness relation\\
Phillips \& Pottasch (1984)& 4.45 & observed vs predicted radio fluxes\\
Maciel (1984)& 1.5 & ionized mass--radius relation\\
Amnuel et al. (1984)& 0.95 & radio surface brightness--radius relation\\
Kingsburgh et al. (1992) & 2.46&  Shklovsky \\
Cahn et al. (1992)& 1.87 & ionized mass--surface brightness relation\\
van de Steene \& Zijlstra (1995) & 1.58 & radio continuum surface brightness temperature--radius relation\\
Zhang (1995) & 1.68& ionized mass--radius and radio cont. surface bright. temp.--radius relations\\
Gorny et al. (1997)& 2.1 & theoretical evolutionary tracks vs observed nebular and stellar parameters\\
Mal'kov (1997) & 2.5 & theoretical evolutionary age vs observed dynamical age\\
Cazetta \& Maciel (2000) & $>$2.9 & Peimbert \& Torres-Peimbert (1983) class$\getsto$stellar mass$\getsto$stellar gravity 
and \\
& &temperature$\getsto$stellar luminosity\\
Bensby \& Lundstr\"om (2001) & 2.23 & ionized mass--radius relation\\
Phillips (2002) & 0.63 & radio surface brightness--radius relation\\
\\
\hline
\end{tabular}
\end{table*}

As compensation the nebula is present in a good 20 catalogues of statistical distances (listed in Table 4). They provide the 
following mean values: 
\begin{description}
\item[]$<$D$>$(Shklovsky)$\simeq$2.0($\pm$0.4) kpc
\item[]$<$D$>$(ionized mass--radius relation)$\simeq$1.7($\pm$0.4) kpc 
\item[]$<$D$>$(surface brightness--radius relation)$\simeq$1.3($\pm$0.4) kpc 
\item[]$<$D$>$(other methods)$\simeq$2.8($\pm$1.0) kpc 
\end {description}

To be noticed that the overall properties inferred in the previous Sections (in synthesis: NGC 6818 is an optically  
thin, almost thin in some directions, quite ``normal'' PN) suggest that $<$D$>$(Shklovsky), 
$<$D$>$(ionized mass--radius relation) and $<$D$>$(surface brightness--radius relation) are, at least grossly, reliable.

In order to derive the dynamical parallax we have analysed the first
and second epoch (1998.30 and 2000.45, respectively) HST images of NGC
6818, searching for the angular expansion of the ionized gas; when
combined with Eq. (1), it provides a reliable nebular distance (Reed
et al. 1999, Palen et al. 2002).  Since the target is at the centre of
the planetary camera (PC) chip in the 2000.45 images, whereas it is
quite off-axis in the 1998.30 ones, the correction for optical camera
distortions was performed with the IRAF/STSDAS task ``drizzle'' (see
Fruchter \& Hook 2002) using the Trauger coefficients.

No apparent shift is obtained from the couples of [O III] and [N II] frames. 

According to Reed et al. (1999), we infer that the 
angular expansion of the gas is $<$1.2$\times$10$^{-2}$ arcsec, that is $\frac{d\theta}{dt}$ $<$5.5$\times$10$^{-3}$ 
arcsec yr$^{-1}$. 
Since D(pc)=0.211[$V{\rm exp}$(km s$^{-1}$)/$\frac{d\theta}{dt}$ (arcsec yr$^{-1}$)], we derive D(NGC 6818)$>$ 1300 pc.

We have attempted a better quantification of D(NGC 6818) through the
interstellar absorption--distance relation (Lutz 1973, Gathier et
al. 1986, Saurer 1995) given by the field stars with accurate m$_{\rm V}$,
m$_{\rm B}$, spectral type and luminosity class. They were selected using
the SIMBAD facilities of the CDS, Strasbourg Astronomical Observatory.
 
The resulting A$_{\rm V}$--distance law is presented in Fig. 8, where the small cluster of data at the right 
edge refers to the super-giant stars of NGC 6822 (Barnard's galaxy), a nearby Ir galaxy projected at about 40 
arcmin S--SE of our nebula.

Fig. 8 indicates that:
\begin{description}
\item[-] the galactic absorption rapidly changes in direction (note the spread of the 
low-weight points). This is confirmed by the inspection of both the ESO/SERC and the Palomar 
Schmidt surveys, showing a variable background crossed by a series of extended, broad and faint emitting filaments;
\item[-] close to the nebula direction (high-weight data), A$_{\rm V}$ quickly increases up to D$\simeq$1.0--1.5 kpc, 
and later is constant (as expected of the large galactic latitude of the field, $|$b II$|\simeq$18$\degr$). Note that 
for the super-giant stars of the nearby galaxy NGC 6822 we have A$_{\rm V}$(tot)=A$_{\rm V}$(galactic)+A$_{\rm V}$(NGC 6822), where 
A$_{\rm V}$(NGC 6822) changes from star to star, and A$_{\rm V}$(galactic)=constant. For A$_{\rm V}$(NGC 6822)=0 we infer 
A$_{\rm V}$(galactic)$\simeq$ 
0.80($\pm$0.1), in excellent agreement with the literature reports (see Gallart et al. 1996, and Bianchi et al. 2001);
\item[-] from A$_{\rm V}$=2.18$\times$c(H$\beta$) (Acker 1978) we have A$_{\rm V}$(NGC 6818)=0.81($\pm$0.06) and D(NGC 6818)
$\ge$1.5 kpc.
\end{description}

Although in Fig. 8 all the solutions for D(NGC 6818)$\ge$1.5 kpc appear legitimate, both the large galactic latitude 
of the nebula ($|$b II$|\simeq$18$\degr$) and the low radial velocity relative to the Local Standard of Rest, 
V$_{\rm LSR}$=-1.7 km s$^{-1}$ (Sect. 4), 
decidedly favour the lowest values, i.e. 1.5 kpc$\le$ D(NGC 6818)$\le$2.0 kpc (in agreement with the information 
previously obtained from the statistical distance scales).

\begin{figure} \centering
\includegraphics[angle=-90,width=9.5cm]{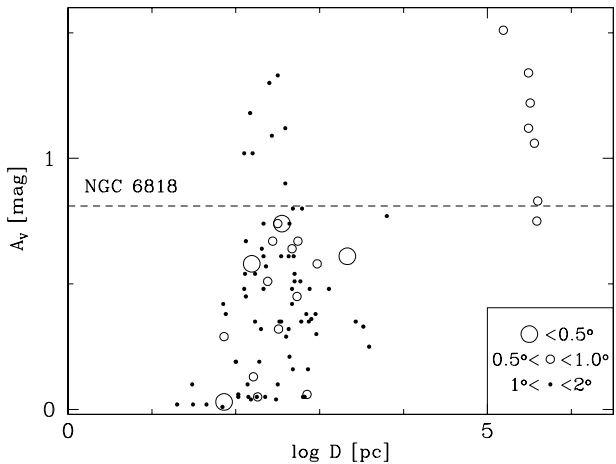} 
\caption{The interstellar absorption--distance relation for the 
field stars of NGC 6818. Three symbols are used, referring to stars at different angular separation from the nebula.}  
\end{figure}

In the following we will adopt D(NGC 6818)= 1.7($^{+0.3}_{-0.2}$) kpc,  
corresponding to a distance from the galactic plane $|z|\simeq$0.5 kpc.

The local filling factor in the nebula is $\epsilon_{\rm l}\simeq$0.5. 
The ionized mass (obtained in different ways: from the H$\beta$ flux, the radio flux and the observed $N{\rm e}$ distribution; 
Aller 1984, Pottasch 1984, Osterbrock 1989 and  
Paper IV) results to be 
M$_{\rm ion}$$\simeq$0.13($\pm$0.03) M$_\odot$, and the kinematical age t$_{\rm kin}\propto$ r/$V{\rm exp}$ $\simeq$ 2300($\pm$300) 
yr. 

M$_{\rm ion}$ is close to the total nebular mass, NGC 6818 being an optically thin (almost thin in 
some directions) PN, whereas t$_{\rm kin}$ represents a lower limit to the actual age, 
$t_{\rm NGC~6818}$, since the dynamical history of the gas is unknown. 
We can obtain a reliable estimate of $t_{\rm NGC~6818}$ by assuming a nebular ejection at 
$V{\rm exp}$$_{\rm superwind}$$<V{\rm exp}$$_{\rm NGC~6818}$, followed by a constant acceleration up to $V{\rm exp}$$_{\rm NGC~6818}$. 
$V{\rm exp}$$_{\rm superwind}$ comes from the OH/IR sources, commonly regarded as the PNe precursors 
(Habing 1996). For $V{\rm exp}$$_{\rm superwind}$ =15($\pm$5) km s$^{-1}$ (Chengalur et al. 1993, David et al. 1993, and 
Sjouwerman et al. 1998) we derive t$_{\rm NGC~6818}$ = 3500($\pm$400) yr, i.e. our nebula is rather young. 

\section{The central star}
As mentioned in Sect. 2, the exciting star of NGC 6818 is the northern component 
of a visual binary system.
The WFPC2 frames taken through the broad-band filter F555W
($\lambda_{\rm c}$= 5407
 $\AA$, bandwidth= 1236 $\AA$) give 
m$_{\rm V}$ = 17.06($\pm0.05$) for the central star and m$_{\rm V}$ = 17.73($\pm0.05$) for the southern red 
companion, where the unknown star colors are the main source of inaccuracy. 
Please note that for D=1.7 kpc we have (M$_{\rm V}$)$_0$(red companion) $\simeq$ +5.75, as expected of a late spectral 
type (G8 to K0) Main Sequence star: a further, although weak, sign in favour of the adopted distance.

According to Tylenda et al. (1993) and Feibelman (1994), the exciting
star of NGC 6818 presents a weak emission line spectrum in both the UV
and the optical regions. In order to derive the H I and He II Zanstra
temperatures, T$_{\rm Z}$H I and T$_{\rm Z}$He II respectively, we have
obtained the total H$\beta$ and $\lambda$4686 $\AA$ nebular fluxes
from the overall line profile at each PA, assuming a circular symmetry
of the spectral image. We find log F(H$\beta$)$_{\rm obs}$=-10.49($\pm0.03$)
mW$\times$m$^{-2}$ and F($\lambda$4686
$\AA$)/F(H$\beta$)=0.65($\pm0.04$), in excellent agreement with the
values reported by Aller \& Czyzak (1979), Kohoutek \& Martin (1981),
Webster (1983), Acker et al. (1991), and Hyung et al. (1999).

The resulting Zanstra temperatures are log(T$_{\rm Z}$H I)= 5.20($\pm0.04$) and log(T$_{\rm Z}$He II)=5.24($\pm0.04$). 
We note in passing that T$_{\rm Z}$H I$\simeq$T$_{\rm Z}$He II; this is quite surprising for a high excitation, optically 
thin (almost thin in some directions) PN. 
Previous determinations (Harman \& Seaton 1966, Martin 1981, Gathier \& Pottasch 1988 and Mal'kov 1997) span the range
4.82 to 5.20 for log(T$_{\rm Z}$H I), and 5.03 to 5.30 for log(T$_{\rm Z}$He II). Moreover Pottasch \& Preite-Martinez 
(1983) and Preite-Martinez et al. (1991) find log T$_*$=4.93 and 5.02--5.05, respectively, from the energy 
balance of the nebula.

The stellar luminosities are log L$_*$/L$_\odot$(T$_{\rm Z}$H I)=3.0 ($\pm0.1$) 
and log L$_*$/L$_\odot$(T$_{\rm Z}$He II)= 3.1($\pm0.1$) (using the bolometric 
corrections by Sch\"onberner 1981, and adopting D=1.7 kpc).

NGC 6818 results to be a rather young PN surrounding a very hot star of relatively low luminosity. 
This suggests that:
\begin{description}
\item[-] the central star is a hydrogen-burning post-AGB star (the helium-burning ones evolve too slowly),
\item[-] the stellar mass, M$_*$, is larger than 
the average value ($\simeq$0.60 M$_\odot$) of the PNe nuclei (Bl\"ocker 1995 and references therein).
\end{description}
The detailed comparison with the theoretical evolutionary tracks by Sch\"onberner (1981, 1983), Iben (1984), 
Wood \& Faulkner (1986), Bl\"ocker \& Sch\"onberner (1990), Vassiliadis \& Wood (1994) and Bl\"ocker (1995) confirms 
that no solutions are possible for the He-burning post-AGB stars (they are too slow), whereas the H-burning nuclei 
give M$_*$=0.625--0.65 M$_\odot$. 

Moreover, the position of the central star of NGC 6818 in the log L--log T diagram coincides with 
the beginning of the luminosity decline at the end of the shell nuclear burning. In this evolutionary phase all 
models present a large luminosity gradient with a deep minimum at log L$_*$/L$_\odot\simeq$3.0, i.e. the value 
obtained for the central star of NGC 6818.
For example the 0.625 M$_\odot$ H-burning nucleus by Bl\"ocker (1995) (showing an astonishing resemblance with 
our star) has log L$_*$/L$_\odot$=3.387 at t=3457 yr, and log L$_*$/L$_\odot$=2.846 at t=3585 yr, i.e. it drops in 
luminosity by a factor of 3.5 in 128 yr.

And indeed, the central star of NGC 6818 is clearly visible in the earliest photographic reproduction of the nebula (Pease 1917; plate 
taken on July 12, 
1912 with the Mount Wilson 60-inch reflector); the rough 
photometric sequence defined by the field stars gives m$_{\rm B}$=15.2($\pm$0.4). 
More stellar magnitude estimates reported in the literature are: m$_{\rm pg}$=14 (Curtis 1918), m$_{\rm pg}$=15 (Berman 1937), 
m$_{\rm pv}$= 15.6 (Cudworth 1974), m$_{\rm V}$$\simeq$14.9 
(Martin 1981), m$_{\rm B}$$>$ 15 (Shaw \& Kaler 1985), m$_{\rm B}$=16.97 and m$_{\rm V}$=17.02 (Gathier \& Pottasch 1988), and 
m$_{\rm V}$=17.06 (this paper, referring to the epoch 1998--2000). 

In addition, the first spectroscopic observation of NGC 6818
(star+nebula, Palmer 1903, referring to 1901) indicates ``a pretty
strong continuum spectrum extending out to $\lambda$372, with the
following bright lines: H$\beta$, H$\gamma$, H$\delta$...''. Palmer
observed a large sample of nebulae with the Crossley reflector+plate
spectrograph, reporting that ``the faintest continuum spectrum
photographed was that of a fifteenth magnitude star''.

In spite of the photometric heterogeneity of these data, the suspicion of a ``historical'' decline for the nucleus 
of NGC 6818 appears legitimate (and robust). 
Also note that the magnitudes of the star just before the luminosity drop were: m$_{\rm V}\simeq$m$_{\rm pv}\simeq$14.5 and  
m$_{\rm B}\simeq$m$_{\rm pg}\simeq$14.0 (for D=1.7 kpc and c(H$\beta$)=0.37).

In summary, all the evidence is that:
\begin{description}
\item[-] the quite massive central star is rapidly fading,
\item[-] the thin-thick transition is hanging over the nebula.
\end{description}

Within this scenario the following caveat is in order: because of the luminosity drop of the fast evolving star, the 
ionization and thermal structure of NGC 6818 are out of equilibrium. The recombination 
processes prevail, being faster in the high ionized species (Tylenda 1986, Marten \& Szczerba 1997); 
this qualitatively explains the observed, unexpected closeness of the H I and He II Zanstra temperatures: the nebula is 
still optically thin (almost thin in some directions), but H$\beta$ maintains a ``longer memory'' than 
$\lambda$4686 $\AA\/$ of the past stellar luminosity. 
Although the large $N{\rm e}$ of the gas implies a short ``time lag'' of the main nebula 
(a few dozen years), nevertheless both T$_{\rm Z}$He II and, even more, T$_{\rm Z}$H I are slightly over-valued. 
In other words: NGC 6818 is over-luminous with respect to the present UV stellar flux.

The overall picture of our nebula shows a remarkable series of analogies with NGC 6565. 
According to 
Paper IV, NGC 6565 is a young (2300 yr), optically thick ellipsoid embedded in a large 
cocoon of neutral, dusty gas. It is in a deep recombination phase (started about 400 yr ago), caused by the 
luminosity drop of the massive powering star (M$_*$$\simeq$0.65M$_\odot$), which has almost reached the white 
dwarf locus (log L$_*$/L$_\odot$$\simeq$2.0, log T$_*$ $\simeq$5.08).

All this suggests the evolutive contiguity of the two PNe, NGC 6818 being a bit older, but also a bit less ``mature'' 
than NGC 6565, i.e. the central star of the latter is more massive. 
No doubt NGC 6818 will develop in the near future most of the observational characteristics already present 
in NGC 6565 (and in the other recombining PNe contained in Tylenda 1986 and Corradi et al. 2000).

We have also carried out a search for other PNe in the peculiar evolutionary phase of NGC 6818 (e.g. at the very 
beginning of the thin-thick transition) adopting the following criteria: high temperature and quite low luminosity of the 
star, 
different [O III] and [N II] nebular morphology, high excitation class and relatively weak low ionization emissions. Among 
the dozen or so candidates we identified, the most promising are NGC 6326, NGC 6884, NGC 7354 and IC 4663.  

\section{The photo-ionization model}
\subsection{General approach}
 \begin{figure*} \centering
\includegraphics[width=17cm]{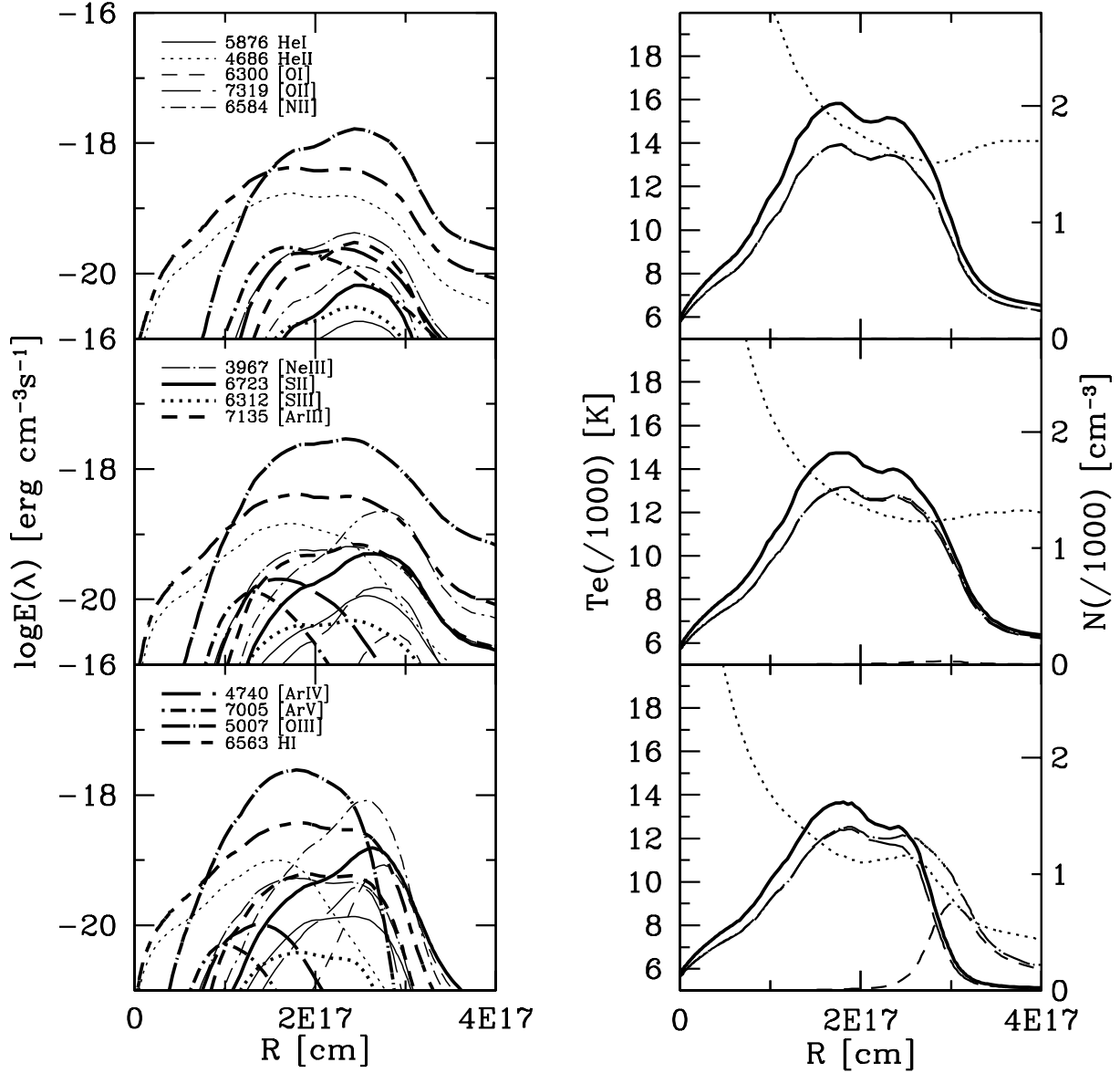} 
\caption{``Steady'' photo-ionization model of NGC 6818, East sector in PA=90$\degr$, at three epochs: (t$_0$-100yr) (upper 
row), t$_0$ (central row), and (t$_0$+100yr) (bottom row), t$_0$=2300 yr being the dynamical nebular age. Left panels: radial 
profiles of the absolute line fluxes. Right panels: the diagnostics; left ordinate scale: $T{\rm e}$ (dotted line); right ordinate scale: 
$N{\rm e}$ (thick continuous line), $N$(H$_{\rm tot}$) (dotted-dashed thin line), $N$(H$^+$) (long-dashed thin line) and 
$N$(H$^0$) (short-dashed thin line). }  
\end{figure*}
In this Section the photo-ionization code CLOUDY (Ferland et al. 1998) is applied to a PN having the same distance, 
matter distribution and chemical composition of NGC 6818.

We immediately point out that CLOUDY is a ``steady-state'' model (a constant UV flux hits the nebula), whereas we argued that 
the ionization and thermal structure of NGC 6818 are out of equilibrium, because of the fast luminosity drop 
of the star. A similar, but less dramatic situation occurs in the recombining PN 
NGC 6565, which is in quasi-equilibrium, since the stellar evolution slows down and the luminosity gradient rapidly decreases 
while approaching the white dwarf domain (Paper IV).

In absence of an ``evolving'' photo-ionization code, which takes into account the gas reactions 
to a fast changing UV flux, at the moment we apply CLOUDY to the present nebula, (NGC 6818)(t$_0$), as well to 
(NGC 6818)(t$_0$-100yr) and (NGC 6818)(t$_0$+100yr). 
Both the past and the future ``snapshots'' come from a homologous expansion 
(i.e. r$\propto$t and $N$(H)$\propto$t$^{-3}$), being  the dynamical time t$_0$= 2300 yr and $\Delta$t$<<$t$_0$. Moreover, the 
evolutionary track of the 0.625 M$_\odot$ 
H-burning post-AGB star by Bl\"ocker (1995) was adopted for the stellar parameters. Last, we have 
selected the quite thin East sector at PA=90$\degr$ as ``mean'' density radial profile of (NGC 6818)(t$_0$).

\begin{table*}
\caption{The CLOUDY photo-ionization code: input parameters of the model nebula}
\begin{tabular}{ll}
\hline
\\
Radial density profile & right panels of Fig.~9 (cfr. Sects. 6.3 and 9)\\
\\
Chemical abundances:   & \\
~~ C, Na, Mg, Si, Cl, K, Ca & Hyung et al. (1999)\\
~~ He, N, O, Ne, S, Ar & this paper\\
~~ other elements      & PN (CLOUDY default)\\
\\
Dust                   & PN (CLOUDY default)\\
\\
Filling factor         & 0.5 \\
\\
                       & at time (t$_0$-100yr): blackbody with T$_*$=170000 K and log L$_*$/L$_\odot$= 3.45 (*)\\
Exciting star         & at time (t$_0$): blackbody with T$_*$=160000 K and log L$_*$/L$_\odot$= 3.10 (see Sect. 8)\\
                       & at time (t$_0$+100yr): blackbody with T$_*$=140000 K and log L$_*$/L$_\odot$= 2.60 (*)\\
\\
& (*) according to Bl\"ocker (1995)\\
\\
\hline
\end{tabular}
\end{table*}

The input data are contained in Table 5. The results, shown in Fig. 9, can be synthesized as follows:
\begin{description}
\item[-] at (t$_0$-100yr) (upper row) the intense UV flux of the luminous and hot star completely ionizes the nebula, which 
appears as 
a very high excitation object at large $T{\rm e}$, optically thin in all the directions;
\item[-] at (t$_0$) (central row) the stellar luminosity is hardly sufficient for ionizing the gas, and the high excitation 
nebula is 
almost thick in some directions, where the low ionization emissions emerge;
\item[-] at (t$_0$+100yr) (bottom row) the UV flux is far inadequate: the innermost plasma maintains 
a high degree of excitation, but the external layers are thick, with prominent [O I], [O II], [N II] and [S II] 
lines. Note that the outermost parts of the ``steady'' photo-ionization model are neutral, whereas 
are recombining in a ``true'' evolving nebula (they create a faint halo embedding the main object). 
\end{description}

Although these results call for caution due to the photo-ionization model limitations, nevertheless they support  
the evolutive scenario of the previous Sections: NGC 6818 is a PN at the very beginning of the recombination phase.

A direct confirmation is expected of the detailed analysis of the
equatorial moustaches, representing the densest and brightest
(i.e. thickest) regions of NGC 6818 (Sect. 5.3). To this end we
have first de-projected the spectral images through Eq. (1), and then
repeated the complete procedure already applied to the zvpc.

\subsection{Equatorial moustaches}

\begin{figure} \centering
\includegraphics[width=9.5cm]{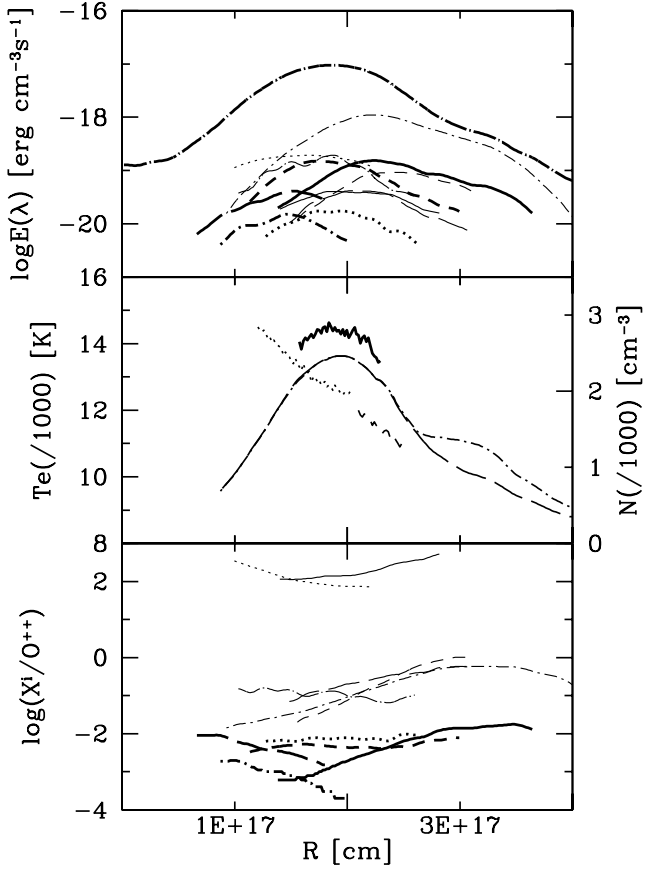} 
\caption{The observed radial properties of the southern, approaching moustache of NGC 6818 in PA=10$\degr$. Top panel: 
the absolute flux (in erg cm$^{-3}$ s$^{-1}$) of the main emissions; same symbols as Figs. 6 and 9. Middle panel: the diagnostics;  
left ordinate scale: dotted line= $T{\rm e}$[O III], short--dashed line= $T{\rm e}$[N II]; right ordinate scale: thick continuous line= 
$N{\rm e}$[S II], 
long--dashed line= $N$(H$^+$) for 
($\epsilon_{\rm l}\times$r$_{\rm cspl}$$\times$D)= 9.5 arcsec kpc, dotted--dashed line= $N$(H$_{\rm tot}$). Bottom panel: the 
$\frac{X^i}{O^{++}}$ ionic abundances (same symbols as the top panel).}  
\end{figure}

Let's consider the southern, approaching moustache in PA=10$\degr$. 
The de-projected radial profiles of the main emissions, the 
diagnostics and the $\frac{X^i}{O^{++}}$ ionic abundances are shown in Fig. 10. 

We are particularly interested to $\lambda$6300 $\AA\/$ of [O I],
marking the neutral (or almost neutral) nebular layers, where the
efficiency of the resonant charge-exchange reaction O$^+$ +
H$^0$$\getsto$O$^0$ + H$^+$ largely increases (Williams 1973).  Note
that on the one hand the [O I] line is very sensitive to the physical
conditions, on the other hand no precise information is yet available
for the electron temperature of the external, neutral (or almost
neutral) gas dominated by the recombination and cooling processes. We
arbitrarily adopt $T{\rm e}$(almost neutral gas)=8000 K, which is lower than
$T{\rm e}$(ionized gas), but not enough low to compromise the [O I]
emissivity (a choice supported by the large kinetic energy of the free
electrons in NGC 6818; see Sect. 5.3).

Moreover, referring to the ionization correcting factor
$icf(O^{++})_{\rm outer}$ of Sect. 6.3, it includes H$^0$, whose
contribution comes from the equilibrium condition O$^+$/O$^0$
$\simeq$0.82$\times($H$^+$/H$^0$), valid for O$^0$$>$O$^+$, i.e. in
the nebular regions affected by the O$^+$ + H$^0$$\getsto$O$^0$ +
H$^+$ reaction (Williams 1973, Stancil et al. 1999).

The observational results of Fig. 10 must be compared with the ``steady'' photo-ionization model of the moustache presented 
in Fig. 11 (the stellar parameters obviously refer to time t$_0$ of Table 5). 

\begin{figure} \centering
\includegraphics[width=9cm]{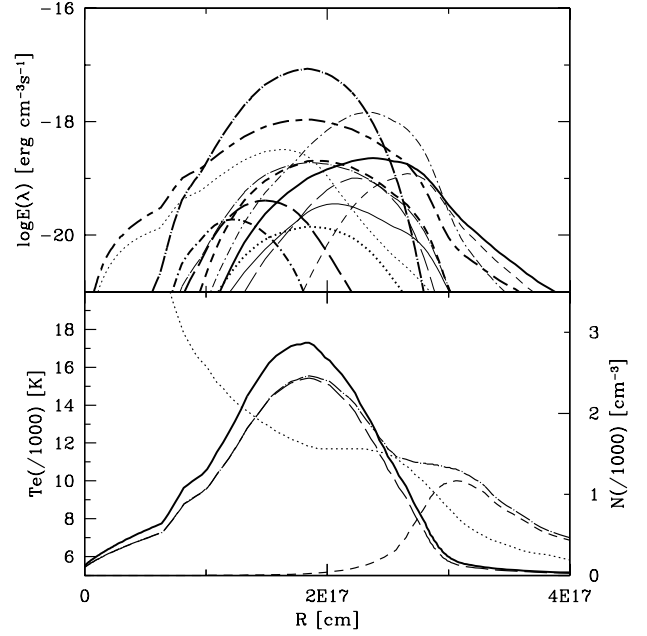} 
\caption{``Steady'' photo-ionization model for the southern, approaching 
moustache of NGC 6818 in PA=10$\degr$. Top panel: absolute radial flux distribution (erg cm$^{-3}$ s$^{-1}$) in the different 
emissions; same symbols as Figs. 6, 9 and 10. Bottom panel: diagnostics radial profile; left ordinate scale: dotted line= 
$T{\rm e}$; right ordinate scale: thick continuous line= $N{\rm e}$, long--dashed line= $N$(H$^+$), short--dashed line= $N$(H$^0$), 
dotted--dashed line= $N$(H$_{\rm tot}$).}  
\end{figure}

Besides the general features common to all the directions (i.e. the decreasing $T{\rm e}$[O III] radial profile, $T{\rm e}$[N II] $< $$T{\rm e}$[O III], 
He$^{++}$/O$^{++}$=He$^+$/O$^{++}\simeq$120, $N{\rm e}$(H$\alpha$)= $N{\rm e}$[S II] for ($\epsilon_{\rm l}\times$ r$_{\rm cspl}$$\times$D$)\simeq$ 
9.5 arcsec kpc, and so on), the moustache of NGC 6818 (Fig. 10) exhibits a remarkable peculiarity: the external 
layers are partially 
neutral, and the outward ionization decline is very smooth, as expected of a recombining region. On the contrary, the 
``steady'' photo-ionization model in Fig. 11 predicts an abrupt ionization fall (to be noticed: the photo-ionization model 
at (t$_0$-100yr), not shown here, indicates that a century ago the moustache was optically thin to the UV stellar flux). 

The $N{\rm e}$ depletion rate for recombination is given by: 
\begin{equation}
dN{\rm e}/dt = -\alpha_{\rm B}\times N{\rm e}\times N(H^+) 
\end{equation}
where $\alpha_{\rm B}$ is the recombination coefficient (Storey \& Hummer 1995). 
Integrating Eq. (13) and assuming $N{\rm e}$=1.15$\times$$N$(H$^+$), we obtain:

\begin{equation}
t =1.15\times\frac{N{\rm e}(t_0-t)-N{\rm e}(t_0)}{\alpha_{\rm B}\times N{\rm e}(t_0-t)\times N{\rm e}(t_0)} 
\end{equation}
which provides the time t elapsed from the beginning of the recombination once 
are known $N{\rm e}(t_0)$, the present electron density, and $N{\rm e}(t_0-t)$, the electron density at the start of the 
process.

The application of Eq. (14) to the external parts of the ``true''
moustache (Fig. 10) furnishes t=30--60 yr (t=60--120 yr and 20--40 yr
for $T{\rm e}$(almost neutral gas)=5000 K and 12000 K, respectively).  In
spite of the heavy assumptions, this agrees with all the previous
evidences suggesting that the recombination phase has just begun in
NGC 6818.

\section{The 3-D morpho-kinematical structure}
\subsection{General} 
The reconstruction of the gas distribution in the nebular
slices covered by the slit was introduced in Papers I and II.
In the case of NGC~6818 
we have selected $\lambda$4686 $\AA$ 
of He II, $\lambda$5007 $\AA\/$ of [O III] and $\lambda$6584 $\AA\/$ of [N II]
as representative of the high, mean and low ionization regions,
respectively. Note that H$\alpha$, the marker of the whole ionized gas
distribution, cannot be utilized because of the blurred appearance.

The spectral images of the forbidden lines have been de-convolved for
seeing, spectral resolution and thermal motions, while also fine
structure has been taken into account for $\lambda$4686 $\AA$ of He
II. They all were de-projected through Eq. (1), and assembled by means
of the 3-D rendering procedure described in Paper III.

In order to reproduce the spatial structure on the paper we adopt the usual method: a series of 
opaque reconstructions of the nebula seen from different directions, separated by 15$\degr$. Each couple forms a stereoscopic 
pair providing a  3-D view of NGC 6818. 

The novelties are represented by the multicolor projection and the movies. 

\subsection{Opaque reconstruction}
\begin{figure*} \centering
\includegraphics[width=7cm]{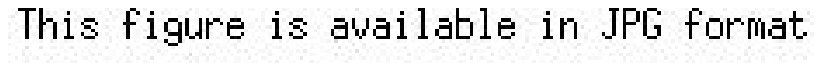} \caption{The structure of NGC~6818 for a rotation around the 
East--West axis centered on the exciting star. Opaque reconstruction at $\lambda$4686 $\AA$
of He II, seen from 13 directions separated by 15$\degr$. The line of view
is given by ($\theta,\psi$), where $\theta$ is the zenith angle and $\psi$ the
azimuthal angle. Each horizontal couple represents a ``direct''
stereoscopic pair, and the whole figure  provides 12 3-D views of the nebula in as
many directions, covering a straight angle. Following 
Paper III: to obtain the three-dimensional
vision, look at a distant object and slowly insert the figure in the field of
view, always maintaining your eyes parallel. Alternatively, you can use the two
small dots in the upper part of the figure as follows: approach the page till
the two dots merge (they appear out of focus); then recede very slowly, always
maintaining the two dots superimposed, till the image appears in focus. The upper-right image is the rebuilt-nebula seen 
from the Earth (West is up and North to the left, to allow the reader the stereo-view).}  
\end{figure*}

For reasons of space we only present the opaque reconstruction in He II, [O III] and [N II] for a rotation of 
180$\degr$ around the East--West axis (close to the minor axis). This is shown in Figs. 12 to 16, 
where the upper--right panel corresponds to 
the nebula seen from the Earth (West is up and North to the left, to allow the reader the stereo-view).

The high excitation layers of NGC 6818 (Fig. 12) form an inhomogeneous shell seen almost equatorial on, open--ended at 
North and South; it merges in a closed ellipsoid at lower $\lambda$4686 $\AA$ fluxes. 

The opaque reconstruction in [O III] is given for two absolute flux cuts: log 
E($\lambda$5007 $\AA$)=-17.32 erg s$^{-1}$ cm$^{-3}$ (Fig. 13), and -17.75 erg s$^{-1}$ cm$^{-3}$ (Fig. 14). 
Since O/H=5.5$\times$10$^{-4}$$\simeq$O$^{++}$/H$^+$ and $N{\rm e}$=1.15$\times$$N$(H$^+$), they correspond to $N{\rm e}$$\simeq$1500 
cm$^{-3}$ (Fig. 13), and $N{\rm e}$$\simeq$900 cm$^{-3}$ (Fig. 14) (for $T{\rm e}$=12000 K and $\epsilon_{\rm l}$=1).  
The [O III] high cut represents the densest regions of the inner shell, mainly constituted by the equatorial moustaches and 
by an extended cup in the southern part, whereas the external, inhomogeneous shell appears in the low cut frames, characterized 
by the large hole at North (along the major axis).
 
The brightest and the faintest [N II] regions refer to $\lambda$6584 $\AA$ at the cuts log 
E($\lambda$6584 $\AA$)=-18.20 erg s$^{-1}$ cm$^{-3}$ (Fig. 15), and -18.80 erg s$^{-1}$ 
cm$^{-3}$ (Fig. 16). The [N II] (high cut)  distribution mimics the corresponding [O III] one, 
whereas at lower fluxes the outermost, knotty structure emerges. Note the general weakness of the low ionization layers for 
-135$\degr<\psi<-45\degr$.

These stereo--reconstructions confirm that NGC 6818 consists of a double shell projected almost equatorial on; the external 
one is spherical (r$\simeq$0.090 pc), 
faint and patchy. It circumscribes a dense and inhomogeneous tri-axial ellipsoid (a/2$\simeq$0.077 pc, a/b$\simeq$1.25, 
b/c$\simeq$1.15) characterized by a large hole along the major axis and a couple of thick equatorial regions (the moustaches).

\begin{figure*} \centering
\includegraphics[width=7cm]{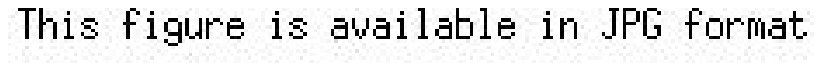} \caption{Same as Fig. 12, but for [O III] at the flux 
cut log 
E($\lambda$5007 $\AA$)=-17.32 erg s$^{-1}$ cm$^{-3}$ ($N{\rm e}$$\simeq$ 1500 cm$^{-3}$ for $T{\rm e}$=12000 K and $\epsilon_{\rm l}$=1).}  
\end{figure*}
\begin{figure*} \centering
\includegraphics[width=7cm]{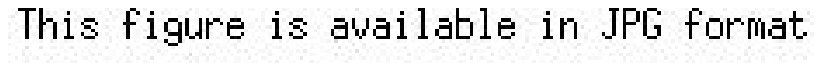} \caption{Same as Fig. 13, but for the [O III] flux cut log 
E($\lambda$5007 $\AA$)=-17.75 erg s$^{-1}$ cm$^{-3}$ ($N{\rm e}$$\simeq$ 900 cm$^{-3}$ for $T{\rm e}$=12000 K and $\epsilon_{\rm l}$=1).}  
\end{figure*}

\subsection{Multicolor projection}
A representative sample of the different morphologies assumed by NGC 6818 when changing the line of view is given in 
Figs. 17 and 18. They show the assembled, multicolor projection at high (He II, blue), 
medium ([O III], green) and low 
([N II], red) ionization for a rotation around the N--S axis (close to the major axis; Fig. 17), and around  
the E--W axis (close to the minor axis; Fig. 18). Also in this case the upper--right panel corresponds to the nebula seen 
from the Earth (North is up and East to the left), to be compared with Fig. 1, and with the HST multicolor image by A. 
Hajian \& Y. Terzian at http://ad.usno.navy.mil/pne/gallery.html. 

These ``almost'' true color reproductions highlight the variety of
looks exhibited by the Little Gem: roundish to elliptical, to
quasi--bipolar (according to the current morphological
classifications; Greig 1972, Stanghellini et al. 1993, Corradi \&
Schwarz 1995, Gorny et al. 1997). NGC 6818 resembles NGC 6153, Hu 1--1
and NGC 4071 when seen from (0,60), K 3--57 and M 2--51 from (0,90), A
70 from (90,0), IC 4663 from (0,30) and NGC 7354 from (150,0) (see the
imagery catalogues of PNe by Acker et al. 1992, Schwarz et al. 1992,
Manchado et al. 1996 and Gorny et al. 1999).
\begin{figure*} \centering
\includegraphics[width=7cm]{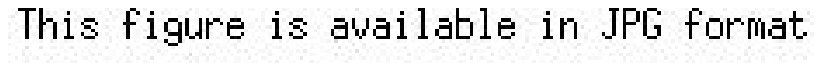} \caption{Same as Figs. 12 to 14, but for [N II] at 
the flux cut log 
E($\lambda$6584 $\AA$)=-18.20 erg s$^{-1}$ cm$^{-3}$ (bright low ionization layers).}  
\end{figure*}
\begin{figure*} \centering
\includegraphics[width=7cm]{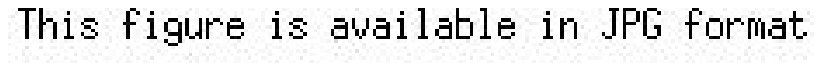} \caption{Same as Fig. 15, but for the [N II] 
flux cut log E($\lambda$6584 $\AA$)=-18.80 erg s$^{-1}$ cm$^{-3}$ (faint low ionization layers).}  
\end{figure*}
\subsection{Movies}

A number of limitations are implicit when showing the 3-D structure on
the paper. They are both objective (space, choice of the rotation
axes, the nebular parameters, the cuts etc.), and subjective
(difficulty in the stereo view). In order to overcome all these
handicaps, we have decided to introduce a series of movies as integral
and functional part of each PN analysis.

Such an ``in fieri'' film library, providing the multicolor
projection, the opaque recovery in different ions and at various cuts,
some slices and radial profiles etc. (suggestions are welcome), can be
found at {\bf http://web.pd.astro.it/sabbadin} (where the acronym
``sabbadin'' stands for ``stratigraphy and best boundary analytic
determination in nebulae'', sic!).

We are also exploring new graphical solutions, rendering at best the
(present) spatial reconstruction and the (forthcoming) spatio-temporal
one. In particular, we are sounding the wide potential of the virtual
reality.

\section{Discussion and concluding remarks} 

In the previous Sections of this paper dedicated to the Little Gem we
have (partially) extracted and interpreted the huge amount of physical
information contained in the ESO NTT+EMMI echellograms by means of the
3-D procedure developed in Papers I to IV for all types of extended,
expanding nebulae.

\begin{figure*} \centering
\includegraphics[width=7cm]{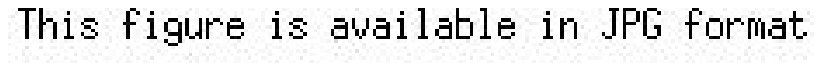} 
\caption{Multicolor appearance of NGC 6818 (blue=He II, green=[O III], 
red=[N II]) for a rotation around the N--S axis centered on the exciting star. The right panel, (0,0), corresponds to the re-built 
nebula seen from the Earth (North is up and East to the left). Same scale as Figs. 12 to 16. Recall that 
projection($\theta,\psi$)=projection($\theta\pm180\degr,\psi\pm180\degr$).}  
\end{figure*}
\begin{figure*} \centering
\includegraphics[width=7cm]{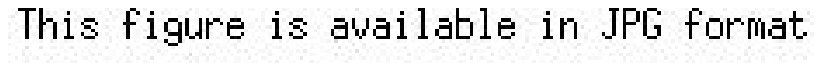} 
\caption{Same as Fig. 17, but for a rotation around the E--W axis.}  
\end{figure*}
NGC 6818 results to be a young (3500 yr), optically thin (quasi--thin
in some directions) double--shell at a distance of 1.7 kpc, projected
almost equatorial--on: a tenuous, patchy spheroid (r$\simeq$0.090 pc)
encircles an inner, dense, inhomogeneous ellipsoid (a/2$\simeq$0.077
pc, a/b$\simeq$1.25, b/c$\simeq$1.15) empty along the major axis and
optically thick in a pair of equatorial moustaches.

We recall that outer attached shells have been found in both the
1D and 2D hydrodynamical models (see e. g. Mellema 1995, Corradi et
al. 2000 and Villaver et al. 2002), due to a D-type ionization front
in the early evolution of the PN. However, all these models fail to
reproduce the smooth, innermost radial density profile observed in NGC
6818 (Fig. 7); they predict an empty region (the hot bubble) and a
quick density rise (the gas compression is provided by the thermal
energy of the hot bubble formed by the adiabatic shock at the
interaction region between the high velocity stellar wind and the
material ejected during the superwind phase).

The central star of NGC 6818 is a visual binary: a faint, red
companion appears at 0.09 arcsec in PA=190$\degr$, corresponding to a
separation $\ge$150 AU. For an orbit of low eccentricity the Kepler's
third law provides a period $\ge$1500 yr.  Note that (by chance?) the
two stellar components appear aligned with the major axis of the
nebula.

Despite some pioneering studies, the physical effects produced by a
wide binary system on the PN ejection and shaping are still poorly
known. Following Soker (1994), an orbital period comparable or longer
than the mass-loss episode generating the nebula causes a density
enhancement in the equatorial plane and/or spiral structures. Soker
(2001) suggests that in wide binary systems (final orbital periods in
the range 40 to 10$^4$ yr) an outer, spherical structure is formed by
the early AGB wind.  Toward the end of the AGB phase, the increased
mass--loss rate creates an accretion disk around the companion. If
this blows jets or a collimated fast wind, two lobes appear in the
inner nebula (a multi--lobed structure in the case of a precessing
accretion disk). Always according to Soker (2001), a fraction of 5$\%$
to 20$\%$ of all PNe originate in such wide binary systems.
Curiously, the same author (Soker 1997) includes (with a high degree
of confidence) NGC 6818 among the PNe resulting from the common
envelope evolution with a sub-stellar companion (planet(s) and/or a
brown dwarf). It is evident that the argument deserves further
attention.

The Little Gem is in a peculiar evolutionary phase, i.e. at the very 
beginning of the recombination process. 
This is caused by the 
luminosity decline of the 0.625 M$_\odot$ central star (log T$_*\simeq$5.22 K; log L$_*$/L$_\odot\simeq$3.1), which has 
recently exhausted the nuclear shell burning and is rapidly moving towards the white dwarf region. 
The stellar drop being 
fated to continue, NGC 6818 will become thicker and thicker, and the amount of neutral, dusty gas in the outermost layers will 
increase with time. The ionization front will re-grow only in some hundreds years, when the gas dilution due to 
the expansion will overcome the slower and slower luminosity decline.

Concerning the observational analogies between NGC 6818 (this paper) and NGC 6565 (Paper IV), 
in Sect. 8 we have pointed out the probable evolutive
contiguity of the two nebulae. We stress here a more facet of the affair, e.g. the importance of the temporal factor: thanks 
to the excellent spatial and spectral resolutions achieved by the 3-D analysis applied to high 
quality spectra, we can no longer regard a PN as a static, uniform and un--changeable object; it is a dynamical, 
inhomogeneous and evolving plasma. 

Ironically, this quite reverses the gap between theory and practice outlined in the 
Introduction: now the existing ``steady'' 
photo--ionization models appear inadequate to interprete the observational data. A 3-D ``evolving'' code is highly desired, 
providing for the gas reactions to the changing UV flux of the central star.

In summary: we have inferred a self-consistent picture of the Little
Gem by means of ESO NTT+EMMI echellograms.  Deeper observations at
even higher spatial and spectral resolutions will disentangle the
still unresolved problems, like the accurate $T{\rm e}$[N II] and $N{\rm e}$[S II]
radial profiles (Sect. 5), the intriguing ionization structure
of the cometary knot in PA=150$\degr$ (Sect. 4) and the possible
blowing of the gas in the Northern and Southern holes (Sect. 4).
Moreover, a gradual change of the [N II]/[O III] morphology is
expected in the future HST imagery, due to the peculiar evolutionary
phase of the nebula.
Concerning the exciting star, a painstaking search in the world-wide
archives (both spectroscopic and photometric), and new, deep, UV to IR
spectra of the stellar system (hot central star + cold companion) are
needed.

At last, the ``vexata quaestio'': which are the mechanisms and the
physical processes ejecting and shaping a PN like NGC 6818?  In our
opinion the question appears premature, and the answer is beyond the
aims of the paper, given the ``forest'' of proposed models (see Icke
et al. 1992, Mellema 1997, Dwarkadas \& Balick 1998, Garcia-Segura et
al. 1999, Frank 1999, Blackman et al. 2001, Soker \& Rappaport 2001,
Balick \& Frank 2002), and the ``desert'' of carefully studied
true nebulae. We are confident that new, reliable and deep insights on
each object and on the whole class will come out of the comparative
analysis of a representative sample of PNe and proto--PNe.

This is the final goal of our survey carried out with  ESO NTT+EMMI and 
TNG+SARG. Indeed, the superb quality of these echellograms constitutes a powerful tool for unveiling the evolutional 
secrets of the PNe (as well for masking the cultural gaps of the authors).

\begin{acknowledgements}
It is a pleasure to thank Gary Ferland, Arsen Hajian, Garrelt Mellema
(the referee), Detlef Sch\"onberner and Noam Soker for their
suggestions, comments, encouragements and criticisms.

This paper has been financied by the grant Cofin MM02905817 of the Italian Ministry of Education (MIUR) and partially 
supported by the grant ASI (Agenzia Spaziale Italiana) I/R/70/00.
\end{acknowledgements}

\end{document}